\documentclass{emulateapj}

\usepackage{color}
\usepackage{ulem}

\newcommand{\el}[2]{\ensuremath{^{#1}\mathrm{#2}}}

\slugcomment{ }

\shorttitle{The $i$ process and CEMP stars}
\shortauthors{M. Hampel et al.}

\begin{document}

\title{The intermediate neutron-capture process and carbon-enhanced metal-poor stars}

\author{Melanie Hampel\altaffilmark{1,2}, Richard J. Stancliffe\altaffilmark{2}, Maria Lugaro\altaffilmark{3,4}, Bradley S. Meyer\altaffilmark{5}}

\altaffiltext{1}{Zentrum f\"ur Astronomie der Universit\"at Heidelberg, Landessternwarte, K\"onigstuhl 12, 69117 Heidelberg, Germany}
\altaffiltext{2}{Argelander-Institut f\"ur Astronomie, University of Bonn, Auf dem H\"ugel 71, 53121 Bonn, Germany}
\altaffiltext{3}{Konkoly Observatory, Research Centre for Astronomy and Earth Sciences, Hungarian Academy of Sciences, H-1121 Budapest, Hungary}
\altaffiltext{4}{Monash Centre for Astrophysics, Monash University, VIC3800, Australia}
\altaffiltext{5}{Department of Physics and Astronomy, Clemson University, Clemson, SC, 29634-0978, USA}
\email{mhampel@lsw.uni-heidelberg.de}

\begin{abstract}

Carbon-enhanced metal-poor (CEMP) stars in the Galactic Halo display enrichments in heavy elements associated with either the $s$ (slow) or the $r$ (rapid) neutron-capture process (e.g., barium and europium respectively), and in some cases they display evidence of both. The 
abundance patterns of these CEMP-$s$/$r$ stars, which show both Ba and Eu enrichment, are 
particularly puzzling since the $s$ and the $r$ processes require neutron densities that are more than ten orders of magnitude apart, and hence are thought to occur in very different stellar sites with very different physical conditions. We 
investigate whether the abundance patterns of CEMP-$s$/$r$ stars can arise from the 
nucleosynthesis of the intermediate neutron-capture process (the $i$ process), which is 
characterised by neutron densities between those of the $s$ and the $r$ processes. Using 
nuclear network calculations, we study neutron capture nucleosynthesis 
at different constant neutron densities $n$ ranging from  
$10^{7}$ to $10^{15}$\,cm$^{-3}$. 
With respect to the classical $s$ process 
resulting from neutron densities on the lowest side of this range, 
neutron densities on the highest side result in abundance patterns that 
show an increased production of heavy $s$-process and $r$-process elements but similar 
abundances of the light $s$-process elements. Such high values of $n$ may occur in the thermal 
pulses of asymptotic giant branch (AGB) stars due to proton ingestion episodes. 
Comparison to the surface abundances of 20 CEMP-$s$/$r$ stars show that 
our modelled $i$-process abundances successfully reproduce observed abundance 
patterns that could not be previously explained by $s$-process nucleosynthesis.
Because the $i$-process models fit the abundances of CEMP-$s$/$r$ stars so well, we propose that this class should be renamed as CEMP-$i$.

\end{abstract}


\section{Introduction}

In the Galactic Halo we can find metal-poor ($\left[ 
\mathrm{Fe}/\mathrm{H}\right]$\footnote{$\left[ \text{A} / \text{B}\right] = \log_{10} 
\left( N_{\text{A}} / N_{\text{B}} \right) _{*} - \log_{10} \left( N_{\text{A}} / 
N_{\text{B}} \right) _{\odot}$, with the number densities $N_{\text{A}}$, $N_{\text{B}}$ 
of element A and B, respectively, where the indices $*$~and~$\odot$ denote the stellar 
and the solar values.}  $< -1$) and very metal-poor ($\left[ 
\mathrm{Fe}/\mathrm{H}\right] < -2$) stars which are amongst the oldest stars that we 
observe. They have formed from almost primordial material that contains the signatures 
of the first nucleosynthesis and chemical enrichment events in the Universe. Multiple 
surveys providing chemical abundances for these low-mass, barely evolved objects 
have revealed a large fraction of carbon-enhanced metal-poor (CEMP) stars 
\citep[e.g.][]{Frebel2006, Lucatello2006, Lee2013, Yong2013, Placco2014}. These are 
generally defined by a carbon excess\footnote{Different authors have adopted other 
definitions of CEMP stars using $\left[ \mathrm{C}/\mathrm{Fe}\right] >0.5$ or $\left[ 
\mathrm{C}/\mathrm{Fe}\right] >0.7$ \citep[e.g.][]{Aoki2007}.} of $\left[ 
\mathrm{C}/\mathrm{Fe}\right] >1$. CEMP stars are subdivided into four classes based on their content of the heavy elements barium 
and europium, which are produced by the slow ($s$) and the rapid ($r$) neutron-capture 
process, respectively. The exact definitions vary 
amongst authors \citep[e.g.][]{Beers2005, Jonsell2006, Masseron2010, Lugaro2012} and the 
categories adopted in this work are:

\begin{itemize}

\item \textbf{CEMP-s} stars show enhancement of barium. This class is defined by $\left[ 
\text{Ba} / \text{Fe}\right] > 1$, $\left[ \text{Ba} / \text{Eu}\right] > 0$ and $\left[ 
\text{Eu} / \text{Fe}\right] \leq 1$ and consequently is thought to show $s$-process 
enrichment only.

\item \textbf{CEMP-$s$/$r$} stars also show barium enhancement, as defined for CEMP-s stars 
with $\left[ \text{Ba} / \text{Fe}\right] > 1$ and $\left[ \text{Ba} / \text{Eu}\right] 
> 0$, but are additionally enriched in europium with $\left[ \text{Eu} / 
\text{Fe}\right] > 1$.

\item \textbf{CEMP-r} stars are enriched in $r$-process elements, in particular in 
europium compared to iron and barium: $\left[ \text{Eu} / \text{Fe}\right] > 1$ and 
$\left[ \text{Ba} / \text{Eu}\right] < 0$. Only a few CEMP stars are currently known 
to fall into this category \citep{Sneden2003, Hansen2015}.

\item \textbf{CEMP-no} stars show no particular enhancements in heavy elements 
\citep{Aoki2002}.

\end{itemize}

The variety of heavy-element abundance patterns observed in CEMP stars points to 
different formation scenarios, in particular due to the differences in the production of 
$s$- and $r$-process elements. The low neutron densities which are required to meet the 
conditions for the $s$ process are approximately \mbox{$n \approx 10^6$\,cm$^{-3}$ to 
$10^{10}$\,cm$^{-3}$} \citep{Busso1999} and result in a neutron capture path which runs 
close to the valley of stability. The predominant producer of $s$-process elements are 
asymptotic giant branch (AGB) stars \citep[e.g.,][]{Gallino1998, Karakas2014}. On the 
other hand, the $r$ process is characterised by neutron densities higher than $\simeq 
10^{20}$\,cm$^{-3}$. Because the $r$ process requires extreme conditions, it is believed 
to occur during supernovae explosions and/or neutron star mergers 
\citep{Thielemann2011, Wehmeyer2015}.

The carbon enhancement of CEMP-no and CEMP-$r$ stars is believed to originate from 
pre-enhancement of the interstellar medium from which these stars formed \citep[e.g.][and 
references therein]{Cooke2014, Frebel2015}. This is supported by observations of very 
metal-poor damped-Lyman $\alpha$ absorption systems that show enrichment in carbon 
\citep{Cooke2011}. These categories of CEMP stars are not discussed further in this 
work. 

The widely accepted origin for the enrichments of carbon and $s$-process elements in 
CEMP-$s$ stars is accretion of matter from an AGB companion in a binary system. Carbon and 
$s$-process elements are known products of AGB nucleosynthesis \citep[e.g.,][]{Karakas2014} 
and studies of radial velocity variations in CEMP-$s$ stars are consistent with all CEMP-$s$ 
stars being in binaries\footnote{\citet{Hansen2016} have called this into question. 
However, the binary star fraction for CEMP-$s$ stars is still significantly higher than 
for other metal-poor stars.} \citep{Lucatello2005, Starkenburg2014}. However, this 
formation scenario cannot explain the origin of CEMP-$s$/$r$ stars because current 
AGB models do not produce enough barium, let alone enough europium \citep{Lugaro2012}. 
Given that we 
believe that the $s$ and the $r$ processes occur in completely different sites under completely 
different conditions, it is puzzling how these stars exhibit signatures of both 
processes. Various formation scenarios have been considered \citep{Jonsell2006, 
Lugaro2009} where the $s$-process enrichment originates from pollution from an AGB 
companion, similarly to the formation scenario for CEMP-$s$ stars. In this framework, 
the additional enrichment in $r$-process elements has been ascribed to: 

\begin{itemize} 
\item a primordial origin, due to pollution of the birth cloud of the binary system by an
$r$-process source \citep{Bisterzo2011} 
\end{itemize} 
or to result from the ejecta of the explosion of 
\begin{itemize} 
\item a third, massive star in a triple system \citep{Cohen2003} 
\item or the primary itself 
\begin{itemize} 
\item either as a type 1.5 supernova \citep{Zijlstra2004, Wanajo2006} 
\item or due to an accretion induced collapse \citep{Qian2003,Cohen2003}. 
\end{itemize} 
\end{itemize} 

\citet{Abate2016} show that 
these formation scenarios have significant difficulties in explaining the observed 
number of CEMP-$s$/$r$ stars, in particular in comparison to the number of CEMP-$s$ and 
CEMP-$r$ stars. Additionally, the observed correlation between the enrichment in $s$- 
and $r$-process 
elements in CEMP-$s$/$r$ stars, as well as high observational ratios of heavy $s$-process 
elements (hs) to light $s$-process elements (ls) provide a challenge 
\citep[e.g.][]{Lugaro2012, Abate2015b, Abate2016}.

As it is difficult to explain the $s$/$r$ abundance pattern via pollution from two independent stellar sites, one may wonder whether it is possible to form both $s$- and $r$-process elements at \textit{the same} site through the action of a modified neutron-capture process operating at neutron densities in between the $s$ and $r$ 
process: the intermediate neutron-capture process (the $i$ process) with 
densities of the order of $n \approx 10^{15}$\,cm$^{-3}$ \citep{Cowan1977}.
Compared to the 
$s$ process, this neutron-capture process should be able to account for both an increased 
production of $r$-process elements as well as a higher hs-to-ls ratio. Initial attempts to explain the abundances of three CEMP-$s$/$r$ stars by the $i$ process have been made by \citet{Dardelet2014}.

For the $i$ process to occur, a neutron burst is required  which is significantly different from that 
occurring in AGB stars responsible for the $s$ process.
There are peculiarities
in the evolution of low-metallicity AGB stars that may allow this to
happen. In AGB stars with low CNO content, the intershell convection
zone that develops during a thermal pulse is able to penetrate up into
the hydrogen burning shell, drawing protons down into hot regions 
\citep[e.g][]{Fujimoto2000, Campbell2008, Lau2009}. These
protons are able to react with the abundant \el{12}{C} present in the intershell
to form \el{13}{C}, which acts as a neutron source via the 
$\el{13}{C}\left( \alpha, \mathrm{n}\right)\el{16}{O}$
reaction. These so-called proton ingestion episodes (PIEs) may 
result in high neutron densities, up to the requisite 
$n =10^{15}$\,cm$^{-3}$ \citep{Cristallo2009b}. A PIE can also develop during
the core helium flash \citep[e.g.][]{Fujimoto1990, Lugaro2009}
with similar consequences for nucleosynthesis \citep{Campbell2010}. More recently, 
calculations by \citet{Jones2016} suggest that PIEs can also take place in the
most massive AGB stars, the super AGB stars \citep[see also][]{Doherty2015}. Unfortunately,
the quantitative predictions of all these 1D stellar evolution
calculations are severely limited by the simplistic treatment of
convection used in these codes. Hydrodynamical simulations of proton
ingestion \citep{Herwig2011, Stancliffe2011, Herwig2014, Woodward2015} show
complex behaviour, the outcome of which is still a matter of debate.

Motivated by the puzzling abundance patterns of CEMP-$s$/$r$ stars and the question of whether 
a single process can explain the signatures of both the $s$ and $r$ process, this study 
investigates the nucleosynthesis of the intermediate ($i$) process. We present 
single zone nuclear network 
calculations under conditions representative of the intershell region of a low-mass, 
low-metallicity AGB star and examine neutron-capture processes at different 
constant neutron densities, to determine whether the abundance patterns in CEMP-$s$/$r$ 
stars can be obtained.

\section{Method}

The nucleosynthesis tools that we use are  \textit{NucNet 
Tools}, a set of C/C++ codes developed by \citet{Meyer2012}.  \textit{NucNet Tools} can 
be used to create nuclear-reaction networks and model the formation of elements in 
stars, supernovae, and related environments. In this study, the codes are used to follow 
the nuclear processing of a single zone with given initial composition under conditions 
of fixed temperature and density. 
One can also specify a species whose abundance should be 
kept constant throughout the calculations, which can be used, for example, to 
artificially induce neutron-capture nucleosynthesis (as we do here) or to imitate a mixing process that 
maintains a constant level of a certain species by mixing in material.

The nuclear network used for this project contains the 5442 isotopes and 45831 reactions 
from the JINA Reaclib V0.5 database \citep{Cyburt2010}. 
In more recent releases, neutron-capture reaction 
rates from KADoNiS v0.2 \citep{Dillmann2006} are included and 
refitted to eliminate blow-ups at low temperatures and to match the 
theory at higher temperatures (JINA Reaclib label $kd02$). However, in some cases the fits  
underpredict the rates in
the temperature regime relevant for the conditions in an AGB intershell region, with 
deviations up to two orders of magnitude, for example, in the case of 
$^{151}$Eu(n,$\gamma$)$^{152}$Eu. Such a large underprediction of the reaction rate introduces 
artificial bottlenecks on the neutron capture path. Because of this, we 
use the previous version of the refitted rates (JINA Reaclib label $ka02$). 
Twenty-eight $\alpha$-decay rates 
\citep{Tuli2011}, which are important for the $s$-process and its termination because of 
being close 
to the valley of stability, were selected and added to the network. These decay rates 
are listed in the online-only Table \ref{tab:alphadecays}.

The physical input conditions are adapted from the density and temperature profiles of 
the intershell region that \citet{Stancliffe2011} found for a low-metallicity AGB star. 
In particular, we present here 
the test case with $T=1.5\times10^{8}$\,K and $\rho = 1600$\,g\,cm$^{-3}$. 
Different temperatures and densities in the range of 
$1.0\times10^{8}$\,K $ \leq T \leq 2.2\times10^{8}$\,K and $800$\,g\,cm$^{-3} \leq \rho 
\leq 3200$\,g\,cm$^{-3}$ were also tested without significant changes in the results.

To model the nucleosynthesis in the intershell region, the composition of the input zone 
is adapted from the intershell composition of \citet[][and references 
therein]{Abate2015}. In particular, we use the abundances of 320 isotopes from 
an AGB star model with metallicity $Z=10^{-4}$ and initial mass $M=1\,M_{\odot}$ after the 
second thermal pulse.

At $n = 10^{15}$\,cm$^{-3}$ the evolution of the abundance distribution is then followed 
for $t=0.1$\,yr, which results in a neutron exposure of 
$\tau=495$\,mb$^{-1}$. The run times of the models at lower neutron densities are scaled 
with $n$ to ensure the same neutron exposure. Such a large value ensures that the 
resulting abundance pattern represents the equilibrium abundance pattern 
between the heavy elements and the seed nuclei. Once this equilibrium is reached, the 
element-to-element ratio is a function of the constant neutron density and is not 
altered by further neutron exposure at the same neutron density. In other words, the abundance 
pattern is independent of the actual neutron exposure, as long as equilibrium is 
reached. 
We note that it is still uncertain what typical neutron exposures are expected from proton ingestion episodes in AGB stars and that realistic values are therefore unknown. To match abundance patterns of CEMP-$s$/$r$ stars with $i$-process nucleosynthesis, \citet{Dardelet2014} found neutron exposures that are about an order of magnitude lower than we assume. However, reducing the neutron exposure by one order of magnitude has a negligible effect on the abundance patterns studied in this work as the heavy elements are already close to equilibrium with one-another: for example, the relative abundance of barium and europium [Ba/Eu] varies by less than $1\%$ during this period.
The limitation of studying equilibrium abundance patterns is that it is not possible to predict abundances at the termination point of the neutron capture path at the lead 
peak. While the element-to-element ratios of the other heavy elements do not change with further neutron exposure, the 
lead-peak elements are only produced and not destroyed by neutron capture processes and 
can therefore not reach an equilibrium. This makes the lead abundance sensitive to 
the final neutron exposure. Adding the total neutron exposure as a degree of freedom should therefore be 
considered in future work to further constrain the $i$ process. Subsequent to the exposure to the 
constant neutron density, the neutron flux is turned off and the successive decays are 
followed for $t=10$\,Myr to allow the long-lived unstable isotopes to decay.

Finally, while we keep the neutron density constant over the whole
time interval and switch the neutron source off instantaneously, a more realistic profile would show the neutron density
decrease with time. We tested this behaviour by
including a smoother decrease and found that the final
abundances are similar to those presented here only if the decrease is
relatively fast with the time scale depending on the neutron density. For
$n = 10^{12}$\,cm$^{-3}$ the decrease can last for about a year before any
changes are seen in the final abundances, while for $n = 10^{15}$\,cm$^{-3}$ the
decrease has to be extremely fast, of the order of a few hours, to keep
the same results as presented here. This effect needs to be further
investigated in the future.

\section{Results and discussion}

\begin{figure}[t!]
\centering
\includegraphics[ width=0.49\textwidth]{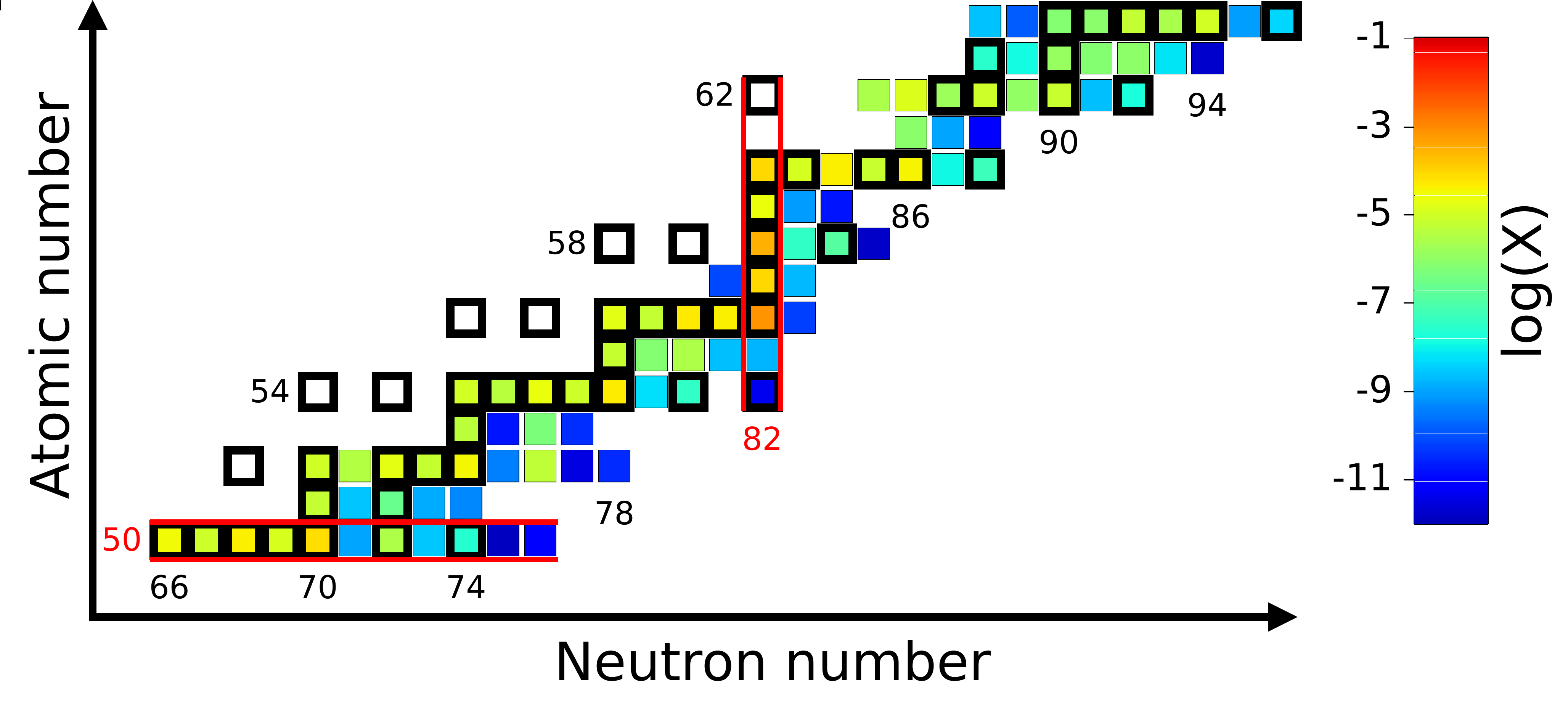} 
\vspace{0.15cm} \\
\includegraphics[ width=0.49\textwidth]{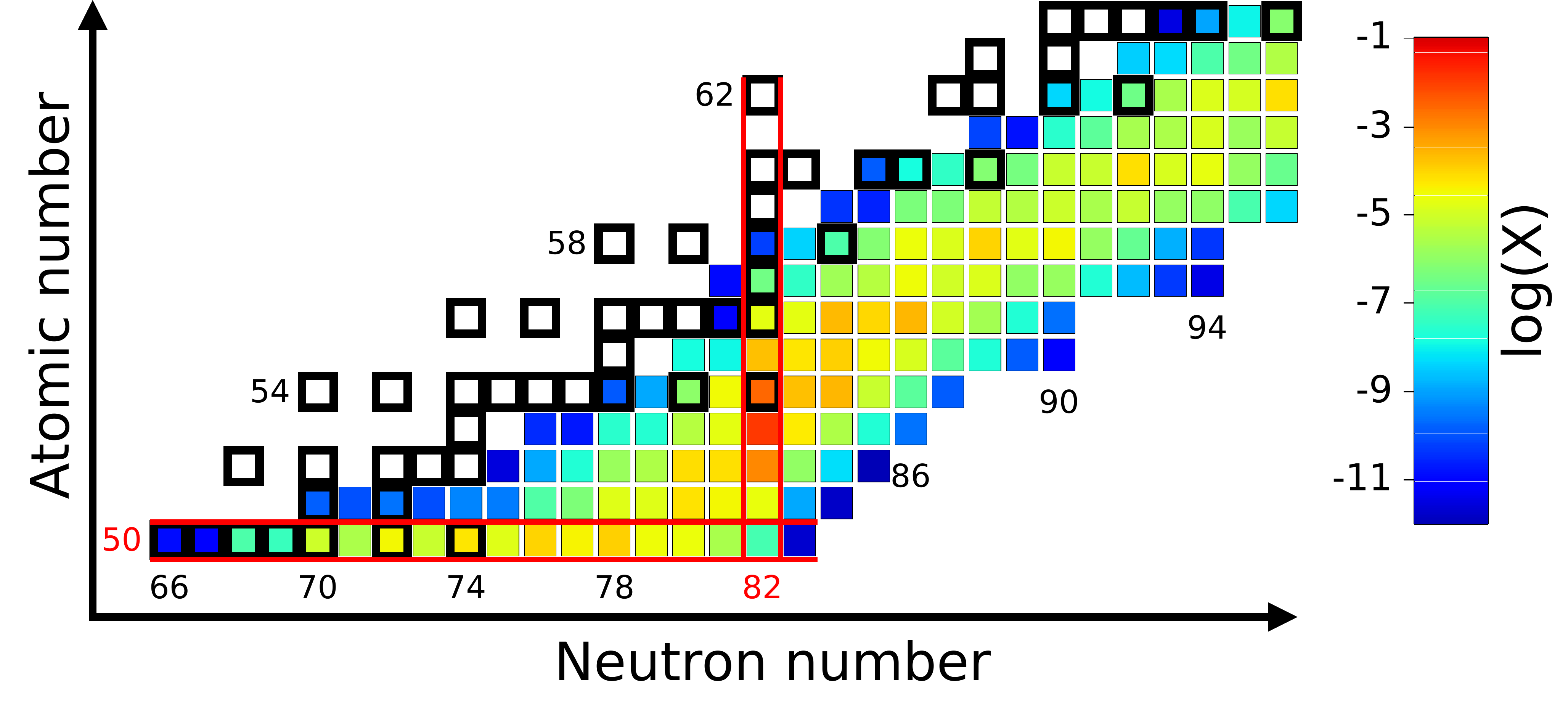} 
\caption{Neutron capture paths of the models with a constant neutron density of $n = 10^{7}$\,cm$^{-3}$ (upper panel) and $n = 10^{15}$\,cm$^{-3}$ (lower panel) shown in the section of the nuclide chart including isotopes of the elements from tin to gadolinium. Isotopes are located as a function of their neutron and proton number and stable isotopes are highlighted by bold black borders. The magic proton and neutron numbers are framed in red. The colours represent the mass fraction of each isotope and thereby show where the neutron capture path produces heavy elements. For $n = 10^{7}$\,cm$^{-3}$ the paths runs mainly through the stable isotopes and stays close to the valley of stability. For $n = 10^{15}$\,cm$^{-3}$ the paths runs much further away on the neutron rich side from the valley of stability. Note the pile-up at $ ^{135} $I with magic neutron number 82 for $n = 10^{15}$\,cm$^{-3}$.}
\label{fig:nuclide_chart}
\end{figure}

When exposed to free neutrons, the present seed nuclei, in particular the abundant iron 
peak nuclei, repeatedly undergo neutron capture reactions. Due to the $\beta$-decays of 
unstable, neutron-rich isotopes, heavy elements are created. Fig. 
\ref{fig:nuclide_chart} compares the neutron-capture paths in a section of the nuclide 
chart for the two different neutron densities of $n = 10^{7}$\,cm$^{-3}$ and $n = 
10^{15}$\,cm$^{-3}$. A higher neutron density creates a neutron capture path further 
away from the valley of stability because an unstable nucleus can form an even heavier 
isotope by neutron capture instead of decaying. Thereby different equilibrium abundance ratios 
between the isotopes are reached that are characteristic for each neutron density. The final heavy-element abundances, i.e. after the decays of unstable nuclei, for the simulations with different neutron densities are listed in detail in the online-only Table \ref{tab:final_abu}.

\begin{figure}[tbh]
\centering
\includegraphics[ width=0.49\textwidth]{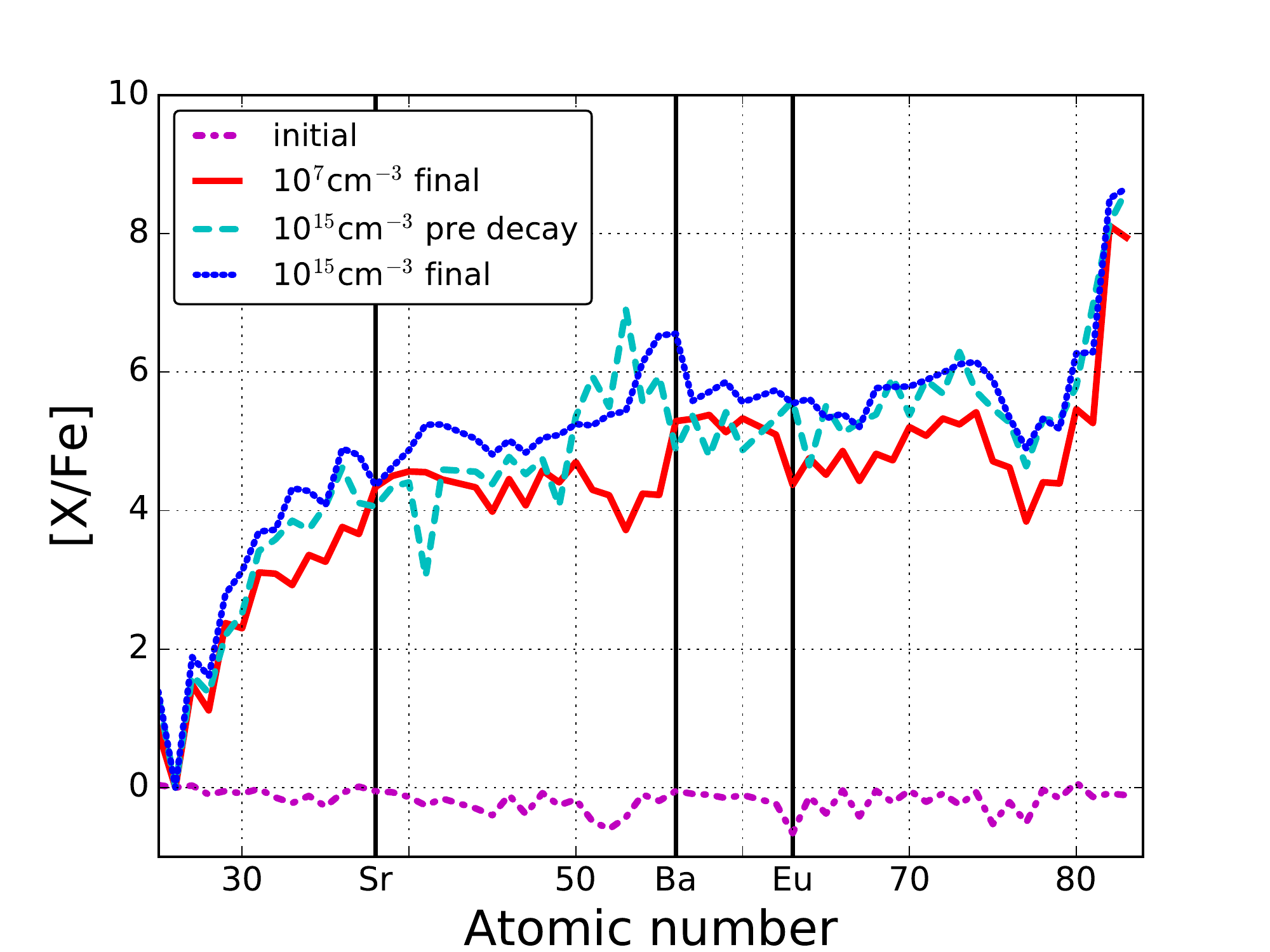} 
\caption{Abundance distributions of the two simulations with neutron densities of $n = 10^{7}$\,cm$^{-3}$ and $n = 10^{15}$\,cm$^{-3}$. For $n = 10^{7}$\,cm$^{-3}$ the pattern created by the active neutron source is sufficiently similar to the final distribution (red line). On the other hand, the nuclei created by the active source with  $n = 10^{15}$\,cm$^{-3}$ are mainly unstable which results in the need to differentiate between the characteristic abundance pattern when the source is active (cyan dashed line) and the final distribution after decays when the source has been switched off (blue dotted line). The initial distribution is shown by the magenta dot-dashed line. The vertical lines show the location of Sr, Ba and Eu which are representatives of the light 
$s$- and heavy $s$-process peak as well as the $r$ process, respectively.}
\label{fig:i-process-summary-XoverFe}
\end{figure}

Fig. \ref{fig:i-process-summary-XoverFe} shows the characteristic equilibrium pattern 
of a neutron source active at $n = 10^{15}$\,cm$^{-3}$ prior to the $\beta$ decays, as well as 
the final heavy-element abundance pattern when unstable nuclei have decayed after the 
neutron source is switched off. For comparison, the resulting abundance pattern of the 
simulation with a neutron density of $n = 10^{7}$\,cm$^{-3}$ is also shown.

For the low neutron density a typical $s$-process abundance pattern arises with 
characteristic elements that are stable bottleneck nuclei with magic neutron numbers. 
These form the light $s$-process (ls) peak (e.g. Sr, Y, and Zr with atomic numbers $Z=38$, 39, and 40, respectively), the heavy $s$-process (hs) 
peak (e.g. Ba, La, Ce with atomic numbers $Z=56$, 57, and 58, respectively), and the lead peak ($Z=82$).

With the active source at $n = 10^{15}$\,cm$^{-3}$, the equilibrium abundance pattern 
shows a shift of the ls and hs peaks to lighter elements, mainly dominated by a peak at 
iodine. This is caused by the shift of the neutron capture path that encounters isotopes 
with magic neutron numbers at lower atomic numbers compared to the $s$ process (compare the panels of Fig. \ref{fig:nuclide_chart}). In particular, the magic, but unstable, isotope 
\el{135}{I} ($Z=53$) acts as the bottleneck of the neutron capture path at $n = 
10^{15}$\,cm$^{-3}$. The final abundance pattern of the $i$ process shows a peak at barium 
due to the decays of the abundant \el{135}{I} into stable \el{135}{Ba}.

The comparison of the final abundance patterns of $n = 10^{7}$\,cm$^{-3}$ and $n = 
10^{15}$\,cm$^{-3}$ shows that the abundance of strontium, a representative of the ls 
peak, does not change. The final abundances of barium and europium, 
representatives of the hs peak and the $r$ process respectively, both increase with neutron density. 
This makes processes at higher neutron densities promising candidates to explain the 
discrepancies found between the CEMP-$s$/$r$ surface abundances and abundance patterns 
modelled with AGB nucleosynthesis simulations by e.g. \citet{Abate2015b, Lugaro2012, Bisterzo2012}.

Furthermore, the distribution of the abundances within
the ls and hs peaks is strongly modified: in the case of the
$s$ process, Sr, Y, and Zr are inevitably overproduced by the same
amount, and the same applies to Ba, La, and Ce. On the other hand, in
the case of the $i$ process, there is an increase of around a factor of 3 between
the production of Sr and that of Zr; and a factor of 10 decrease
between Ba and La and/or Ce. In fact, the overproduction of Ba with respect
to La resulting from the $i$ process has been pointed out as the
possible source of the Ba excess observed in open clusters
\citep{Mishenina2015}.

\subsection{Comparison to CEMP-$s$/$r$ stars}

The abundances of 67 CEMP stars with barium enhancement were studied by \citet{Abate2015b}, 20 of which are classified as CEMP-$s$/$r$ stars. This sample of CEMP-$s$/$r$ stars is based on the SAGA database \citep{Suda2008}. The objects and their properties are listed in Table \ref{tab:data_sample} and were selected in the metallicity range $-2.8 \leq \left[ \text{Fe}/\text{H}\right] \leq -1.8$. For further information regarding the data sample see \citet{Abate2015b}\footnote{For HE1305+007 information on Zr is available from \citet{Goswami2006}. We add $\left[ \text{Zr}/\text{Fe}\right]=2.09 \pm 0.3$ to the data used by \citet{Abate2015b}.}. As \citet{Abate2015b} were unable to obtain satisfactory fits to these 20 stars using standard $s$-process calculations, we shall compare the final abundance patterns of our $i$-process calculations with neutron densities of $n = 10^{12}$, $10^{13}$, $10^{14}$, and $10^{15}$ \,cm$^{-3}$ to these objects to see if we can obtain better fits.

To compare the model abundances to the surface abundances of the CEMP-$s$/$r$ stars, we need to mix an unknown quantity of $i$-processed material with non-processed material. Some mixing will occur because dredge-up must presumably extract the $i$-processed material from the stellar interior to its envelope. As we assume that CEMP-$s/r$ stars form in a similar manner to CEMP-$s$ stars (i.e. from mass transfer in binary systems), further dilution can also occur once this material is transferred to its companion \citep{Stancliffe2007, Stancliffe2013}. We therefore assume that the convective envelope and the secondary star both
have initially Solar element-to-element ratios for the heavy elements,
although it should be kept in mind that this initial relative
composition could have been different from Solar, and closer to that
of pristine Galactic material, such as observed in CEMP-no stars. In
any case, because these stars do not show enhancements in elements
heavier than iron, the final resulting pattern would not be greatly
affected. The modelled surface abundances of the CEMP-$s$/$r$ star can then be computed using the following equation:
\begin{eqnarray*}
X = X_i \times \left( 1-d\right) + X_{\odot} \times d\, , 
\end{eqnarray*} 
where $X$ is the final abundance, $X_i$ the abundance from the modelled $i$ process after the decays, $X_{\odot}$ the solar-scaled abundance and $d$ a dilution factor. The dilution factor is a free parameter in these simulations and is varied in order to find the best fitting model to the observational data for each simulation with a different constant neutron density. In order to find the best fitting model $\chi^2$ is computed for each simulation from:
\begin{eqnarray*}
\chi^2 = \sum_{Z} \frac{\left( \left[ X_{Z}/\text{Fe}\right]_{\text{obs}} - \left[ X_{Z}/\text{Fe}\right]_{\text{mod}}\right) ^2}{\sigma_{Z,\text{obs}}^2} \, ,
\end{eqnarray*}
where $\left[ X_{Z}/\text{Fe}\right]_{\text{obs}}$ and $\left[ X_{Z}/\text{Fe}\right]_{\text{mod}}$ are the observed and modelled abundances, respectively, of the element with atomic number $Z$ and $\sigma_{Z,\text{obs}}$ is the observational error of $\left[ X_{Z}/\text{Fe}\right]_{\text{obs}}$. For these calculation, the abundances of the heavy elements with $30 < Z \leq 80$ were taken into account. The abundances of elements with $Z\leq30$ are not significantly produced in the studied neutron-capture processes and are therefore not considered. We explicitly exclude Pb from the computation, because the final Pb abundance depends on the actual neutron exposure. For all the other elements, once equilibrium is achieved during neutron exposure, the abundance ratios will not change. To quantify the deviations between the predictions of the best-fitting model and the observed abundances, the residual is defined as
\begin{eqnarray}
R_Z = \left[ X_{Z}/\text{Fe}\right]_{\text{obs}} - \left[ X_{Z}/\text{Fe}\right]_{\text{mod}} \, .
\label{eq:residual}
\end{eqnarray}

\begin{figure}[tbhp]
\centering
\includegraphics[width=0.45\textwidth]{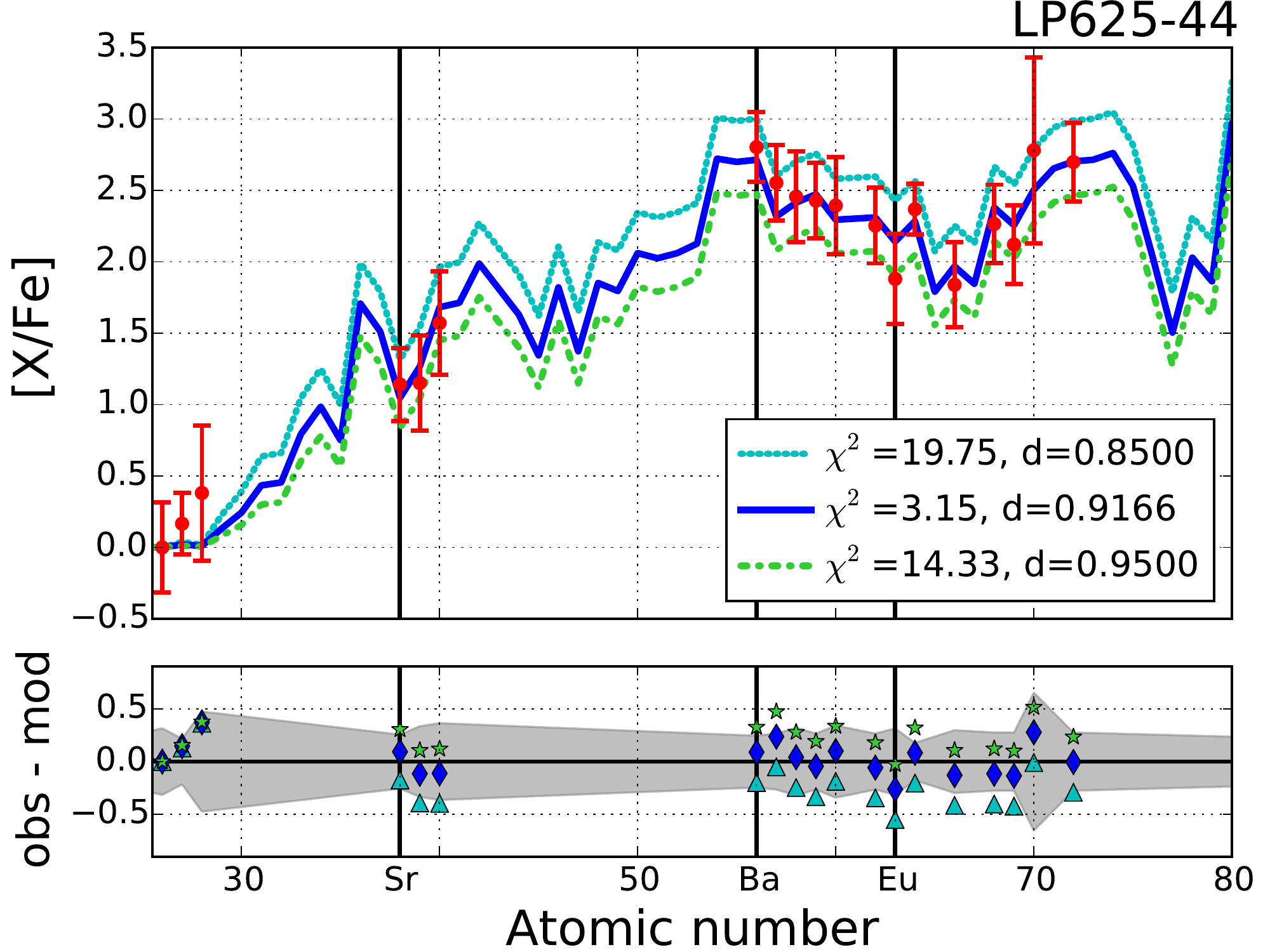}
\caption{Influence of the dilution factor on the modelled abundances of CEMP-$s$/$r$ 
star LP625-44. The observed surface abundances (red dots) and the best-fitting model at 
$n = 10^{14}$\,cm$^{-3}$ with a dilution factor of $d=0.917$ (blue) are compared to the 
abundance distributions with two alternative dilution factors of $d=0.850$ (cyan) and 
$d=0.950$ (green). The lower panel shows the distribution of the residuals as defined in 
Equation \ref{eq:residual}. The uncertainty of the observations $\sigma_{Z,\text{obs}}$ 
is indicated by errorbars in the upper panel and by the shaded region in the lower 
panel. The vertical lines show the location of Sr, Ba and Eu which are representatives 
of the ls and hs peak as well as the $r$ process, respectively.}
\label{fig:vary_d}
\end{figure}

The following example fit shows the influence of the two degrees of freedom in the fits, 
namely the dilution factor, $d$, and the neutron density, $n$. Fig. \ref{fig:vary_d} 
shows the abundance pattern of the CEMP-$s$/$r$ star LP625-44 and the best fitting model 
with a neutron density of $n = 10^{14}$\,cm$^{-3}$ and a dilution factor $d=0.917$. For 
comparison, two alternative models with a higher and a lower dilution factor at the same 
neutron density are shown. The higher the dilution factor, the lower are the abundances 
of heavy elements. Therefore the main influence of the dilution factor is to set the 
right abundance level to match that of the CEMP-$s$/$r$ star. Table \ref{tab:vary_n} lists 
the best-fitting dilution factors for each of the tested neutron densities along with 
the minimal $\chi^2$ for each $n$. While $d$ predominantly determines the overall 
abundance level, the variation of $n$ has a higher impact on the individual 
element-to-element ratios and hence influences the quality of the fit indicated by the 
value of $\chi^2$. We note that for some systems (e.g., LP625-44), the reduced $\chi^2$ is less than 1. While this could be interpreted as suggesting the error bars are too large, one should not rely on the statistical quantitative meaning of $\chi^2$ because the errors on abundance measurements are not simply Gaussian measurement errors. Therefore $\chi^2$ is just an indicator to identify which model matches the observations best, but its value does not have the statistical meaning in the usual sense.

\begin{table}[tbhp]
  \centering
      \caption{Fit parameters for the CEMP-$s$/$r$ star LP625-44. The best-fitting model is the one with $n = 10^{14}$\,cm$^{-3}$, as indicated by a clear minimum in $\chi^2$.}
    \begin{tabular}{c|c|c}
    \toprule
$n$ (cm$^{-3}$) & $d$ & $\chi^2$ \\
\tableline
$10^{12}$ & 0.932 & 8.2 \\
$10^{13}$ & 0.933 & 6.4 \\
$10^{14}$ & 0.917 & 3.2 \\
$10^{15}$ & 0.862 & 8.7 \\
\tableline
    \end{tabular}%
  \label{tab:vary_n}%
\end{table}%

The details of the best-fitting model for all 20 stars, i.e., how many measurements the 
fit is based on, the neutron density of the model, dilution factor and $\chi^2_{min}$ of the 
best fit, are listed in Table \ref{tab:fit_summary}. 
For most of the stars the majority of the elemental abundances can be reproduced within 
the uncertainty of the observational measurements. 
One significant exception is SDSSJ0912+0216 which has an unusual abundance pattern that is unlike the other stars in the sample. It cannot be reproduced by either an $i$ or $s$ process. Further study of this object is warranted.
Two stars, BS16080-175 and BS17436-058, only have measurements of heavy elements for barium, lanthanum 
and europium that the fit can be based on. Due to this low number of observations, their fits are 
less meaningful than for the remaining stars with significantly more observed abundances.
Interestingly, most of the abundance patterns of the remaining 17 stars can be best modelled by a 
neutron capture process operating at a neutron density of $n = 10^{14}$\,cm$^{-3}$, which is the case 
for 12 stars. Four stars are better described by processes operating at the lower neutron 
densities of $n = 10^{12}$\,cm$^{-3}$ (CS22881-036 and HD187861) and $n = 10^{13}$\,cm$^{-3}$ 
(CS22948-027 and HD224959). The only star for which the best fit to the data is achieved by the model 
of the $i$ process operating at a neutron density of $n = 10^{15}$\,cm$^{-3}$ is CS31062-050. 
However, the abundances of CS31062-050 can be modelled almost as well by the simulation of $n = 
10^{14}$\,cm$^{-3}$ with $\chi^2=26.7$ compared to $\chi^2=26.5$ for $n = 10^{15}$\,cm$^{-3}$. 
Therefore it is arguable that a neutron density around $n = 
10^{14}$\,cm$^{-3}$ is sufficient to reproduce the abundance patterns of most CEMP-$s$/$r$ stars, because 
this is the case that results in both high Eu abundances and 
$\left[\text{Ba}/\text{Eu}\right]\approx0.6$ as observed in CEMP-$s$/$r$ stars.

\subsection{Comparison to other studies}

While the original idea for the $i$ process is not new \citep{Cowan1977}, there have been 
few studies of its production of the heavy elements to which we can compare our results. 
In the context of CEMP-$s$/$r$ stars, 
\citet{Dardelet2014} examined its effects on three CEMP-$s$/$r$ stars:
CS22898-027, CS31062-050, HE0338-3945. Like our simulations, 
these authors used a single-zone nucleosynthesis code to compute the effects of the 
$i$ process but rather than using a constant neutron density, they adopt a constant combined 
C+H mass fraction of $0.7$ to simulate proton ingestion. For the three systems they 
studied they found similar fitting neutron 
densities to those we have obtained and essentially the same resulting abundance pattern.

\begin{figure}[tbh]
\centering
\includegraphics[width=0.45\textwidth]{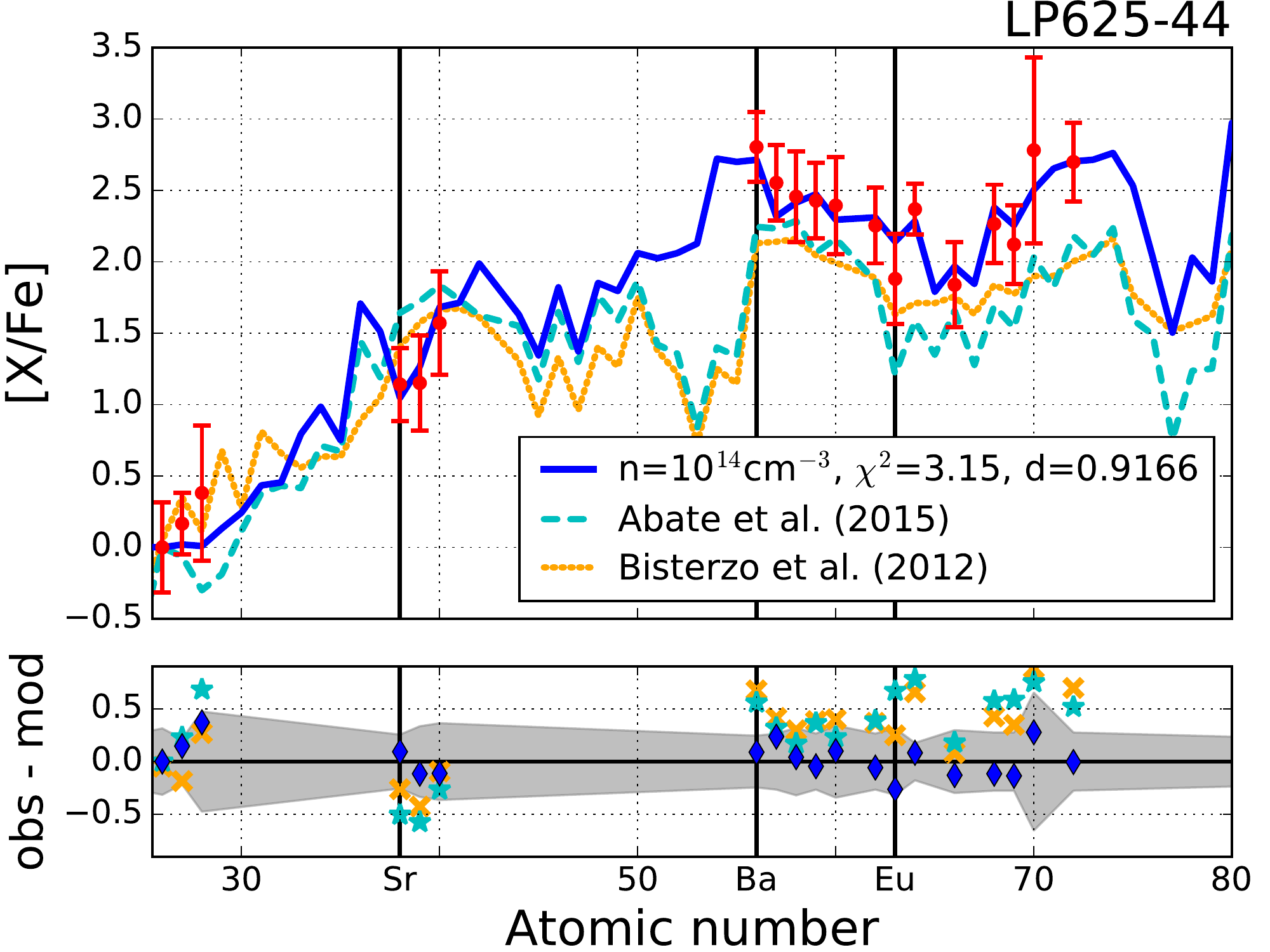}
\caption{Best fitting model for CEMP-$s$/$r$ star LP625-44 (red dots): the best-fitting models from \citet{Abate2015b} with AGB nucleosynthesis (cyan) and from \citet{Bisterzo2012} with s-process and initial [r/Fe]=1.5 (orange) compared to the best-fitting model from the neutron capture nucleosynthesis calculations with a neutron density of $n = 10^{14}$\,cm$^{-3}$ (blue). Lower panel, vertical lines and uncertainties as in Fig. \ref{fig:vary_d}.}
\label{fig:cp_carlo_star}
\end{figure}

Fig. \ref{fig:cp_carlo_star} shows the observed abundance pattern of LP625-44 and the 
best-fitting model from this work, along with the best fitting model from the studies 
of \citet{Abate2015b} and \citet{Bisterzo2012}. It can be seen that 
the main problems of the fit with standard AGB nucleosynthesis - 
to explain the high [hs/ls] ratio as well as the high Eu ($Z=63$) abundance - 
are almost entirely resolved 
by modelling the CEMP-$s$/$r$ surface abundances with $i$-process neutron-capture 
nucleosynthesis with $n = 10^{14}$\,cm$^{-3}$.
Best fit $s$-process models present further problems in reproducing the 
abundances of the elements between Eu and Hf ($Z=72$) and relative abundance variations within the first and second $s$-process peaks, where Zr and Ba are often observed to be higher than Sr--Y and La--Ce, 
respectively. 

Because of these problems, \citet{Bisterzo2012} considered diluting $s$-processes material not with pristine material, but with matter that was pre-enriched in $r$-process elements. Diluting $s$-process material with $r$-processed material presents similar problems to those described above, even when the initial $r$-process abundances are assumed to be enhanced by [r/Fe]=1-2 dex 
in order to match the [Eu/Fe] abundance (see, e.g., the best fit for LP625-44 in Fig.~31 of \cite{Bisterzo2012}). 
Patterns which match much better are instead found in our $i$-process results. 

\begin{figure}[tbh]
\centering
\includegraphics[width=0.45\textwidth]{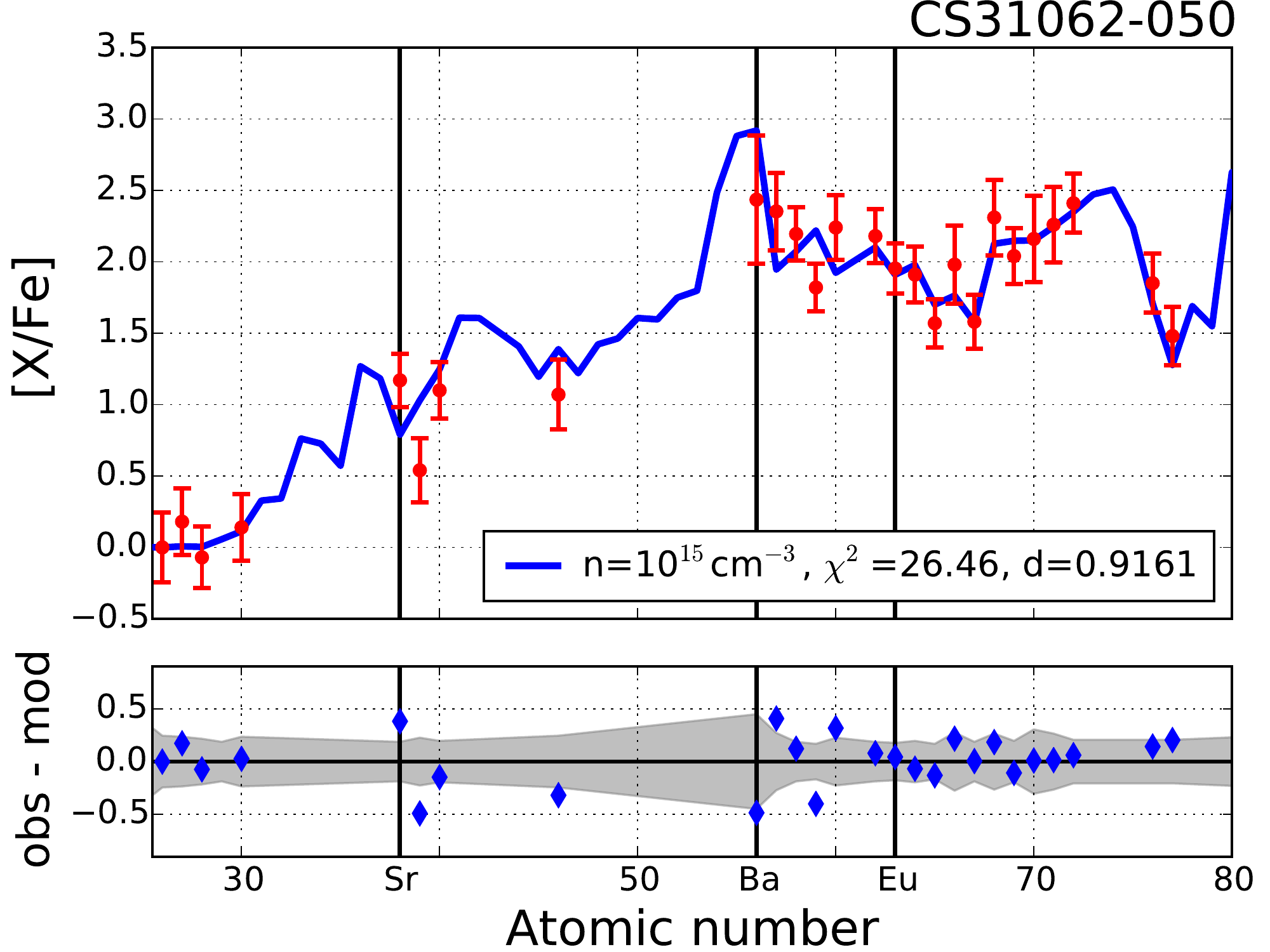}
\caption{Best fitting $i$-process model for CEMP-$s$/$r$ star CS31062-050 (red dots).
The $s$-process best fit with initial [r/Fe]=1.6 can be found in Fig.~26 of \citet{Bisterzo2012}.}
\label{fig:CS31062-050}
\end{figure}

\begin{figure}[tbh]
\centering
\includegraphics[width=0.45\textwidth]{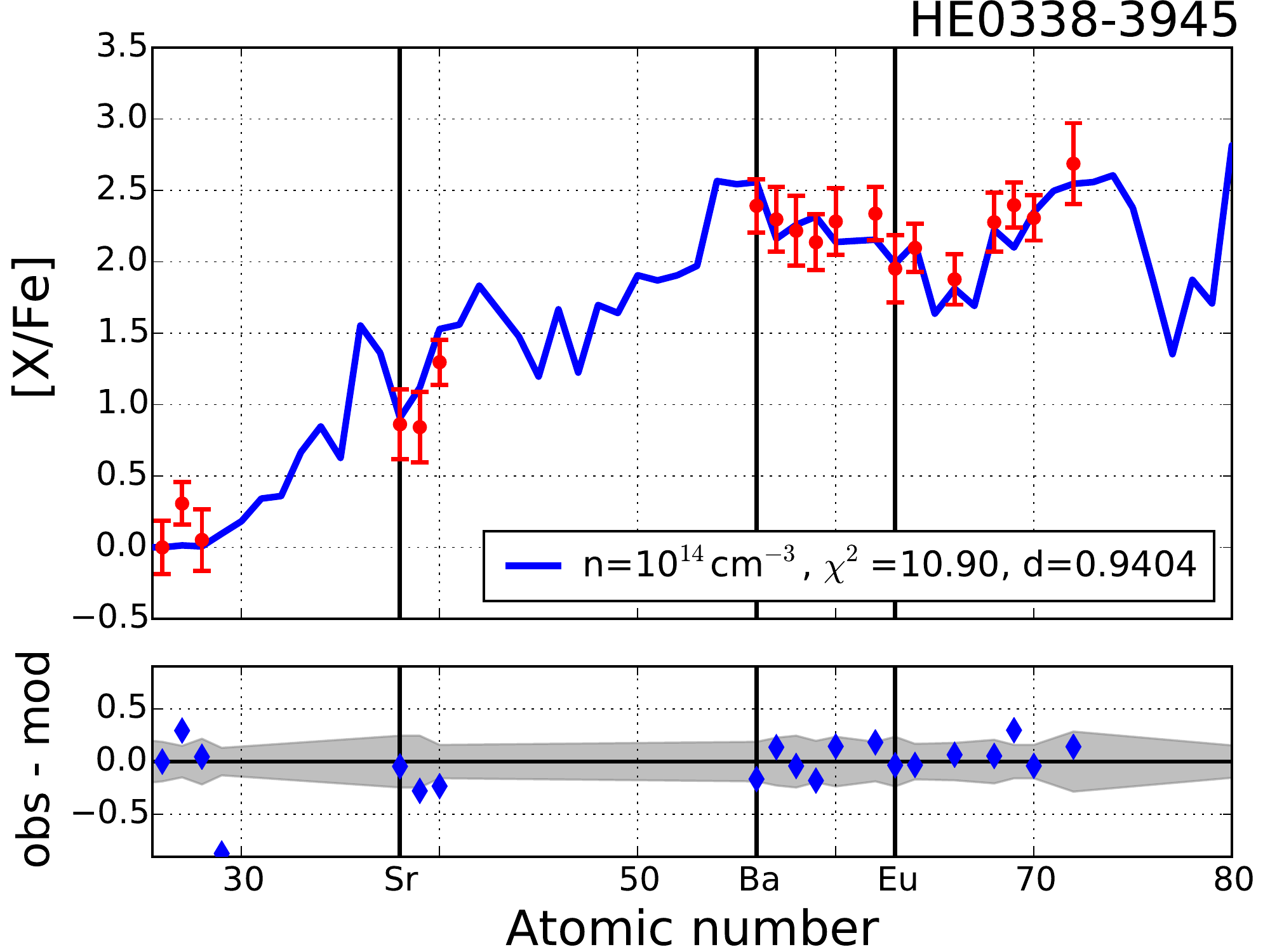}
\caption{Best fitting $i$-process model for CEMP-$s$/$r$ star HE0338-3945 (red dots).
The $s$-process best fit with initial [r/Fe]=2 can be found in Fig.~19 of \citet{Bisterzo2012}.}
\label{fig:HE0338-3945}
\end{figure}

\begin{figure}[tbh]
\centering
\includegraphics[width=0.45\textwidth]{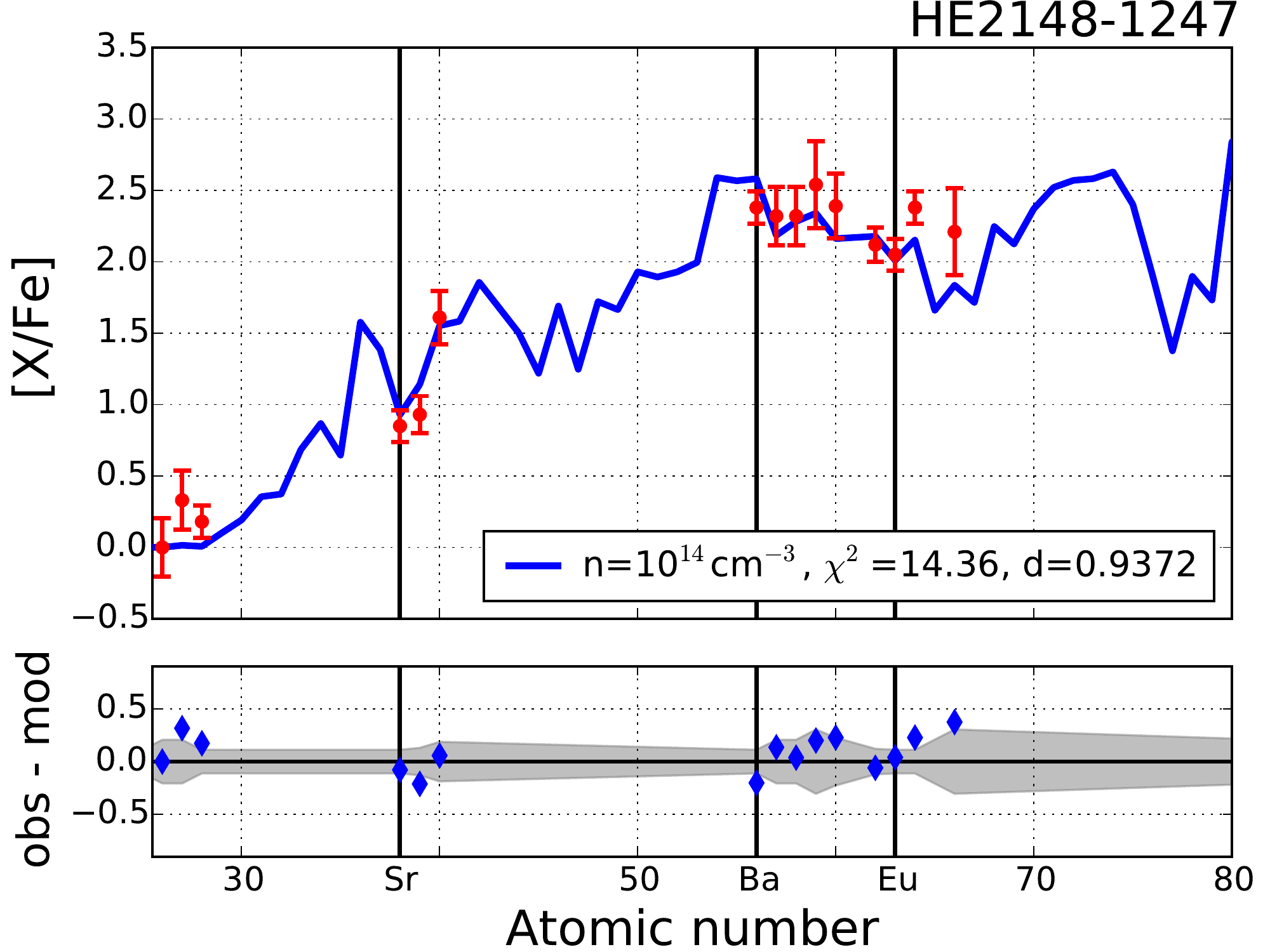}
\caption{Best fitting $i$-process model for CEMP-$s$/$r$ star HE2148-1247 (red dots).
The $s$-process best fit with initial [r/Fe]=2 can be found in Fig.~21 of \citet{Bisterzo2012}.}
\label{fig:HE2148-1247}
\end{figure}

In Figs.~\ref{fig:CS31062-050} to 
\ref{fig:HE2148-1247}, we show further examples of our $i$-process fits and the captions provide references to the fits of \citet{Bisterzo2012} for comparison.
Our remaining individual fits can be found in Appendix~\ref{sec:all_fits}. The stars presented in Figs.~\ref{fig:cp_carlo_star} to \ref{fig:HE2148-1247} have the largest number of observed elements (between 25 and 37), providing the 
most stringent test bed for our model. Detailed inspection of these figures 
demonstrates that there is a an excellent match between our models and the
observations, superior to any $s$-process fits presented so far. 
Furthermore, the match not only involves the abundance of Eu, but also 
several other relatively minor details that have been problematic for the 
$s$-process models: the relative abundances of Sr, Y, Zr and of Ba, La, and 
Ce; the elements between Eu and Hf; the abundance of Pd ($Z=46$), as discussed 
below. For example, a clearly better match with the abundances of the elements between Eu and Hf is shown by 
LP625-44 (Fig.~\ref{fig:cp_carlo_star}) and CS31062-050 (Fig.~\ref{fig:CS31062-050}). A match with the observed 
positive [Ba/La] and [Ba/Ce] ratios is shown by LP625-44 (Fig.~\ref{fig:cp_carlo_star}), CS31062-050 
(Fig.~\ref{fig:CS31062-050}) and HE0338-3945 (Fig.~\ref{fig:HE0338-3945}). It is also possible to obtain a 
better match with the abundance pattern of the first $s$-process peak as shown by all the plotted 
individual fits, although in most cases (for example CS31062-050, Fig.~\ref{fig:CS31062-050}) it appears 
that a shift of the local abundance minimum from Sr to Y would provide a better match. This issue may be 
related to nuclear uncertainties in the production of the first peak. Finally, the best $s$-process fits for CS31062-050 show an 
overabundance of Pd, which required \citet{Bisterzo2012} to assume a further ``light-element $r$-process enhancement'' of 0.5 dex for the 
elements from Mo ($Z=42$) to Cs ($Z=55$). The $i$ process on the other hand naturally explains an increase in Pd (Fig.~\ref{fig:CS31062-050}), even to an extent that slightly overpredicts the measurement.

\begin{table}[htbp]
  \centering
      \caption{Fit parameters for each CEMP-$s$/$r$ star: number of measurements the fit is based on, neutron density $n$, dilution factor $d$ and minimum $\chi^2$.}
    \begin{tabular}{ccccc}
    \toprule
    ID & $N_{\text{obs}}$ & $\log \left( n/\text{cm}^{-3}\right) $ & $d$ & $\chi^2_{min}$ \\
    & $\left( 31\leq Z \leq 80\right)$ & & & \\
\tableline
BS16080-175 & 3 & 12 & 0.991 & 2.0 \\
BS17436-058 & 3 & 13 & 0.989 & 0.2 \\
CS22881-036 & 7 & 12 & 0.985 & 5.1 \\
CS22898-027 & 11 & 14 & 0.937 & 5.7 \\
CS22948-027 & 9 & 13 & 0.965 & 5.9 \\
CS29497-030 & 15 & 14 & 0.957 & 8.1 \\
CS29526-110 & 7 & 14 & 0.966 & 2.7 \\
CS31062-012 & 7 & 14 & 0.971 & 3.2 \\
CS31062-050 & 22 & 15 & 0.916 & 26.5 \\
HD187861 & 8 & 12 & 0.978 & 0.5 \\
HD224959 & 8 & 13 & 0.969 & 3.7 \\
HE0131-3953 & 6 & 14 & 0.969 & 0.4 \\
HE0143-0441 & 8 & 14 & 0.947 & 9.0 \\
HE0338-3945 & 16 & 14 & 0.940 & 10.9 \\
HE1105+0027 & 6 & 14 & 0.953 & 1.2 \\
HE1305+0007 & 10 & 14 & 0.858 & 7.6 \\
HE2148-1247 & 12 & 14 & 0.937 & 14.4 \\
HE2258-6358 & 17 & 14 & 0.973 & 23.4 \\
LP625-44 & 16 & 14 & 0.917 & 3.2 \\
SDSSJ0912+0216 & 16 & 14 & 0.938 & 373.7 \\
\tableline
    \end{tabular}%
  \label{tab:fit_summary}%
\end{table}%

\begin{figure}[tbh]
\centering
\includegraphics[width=0.45\textwidth]{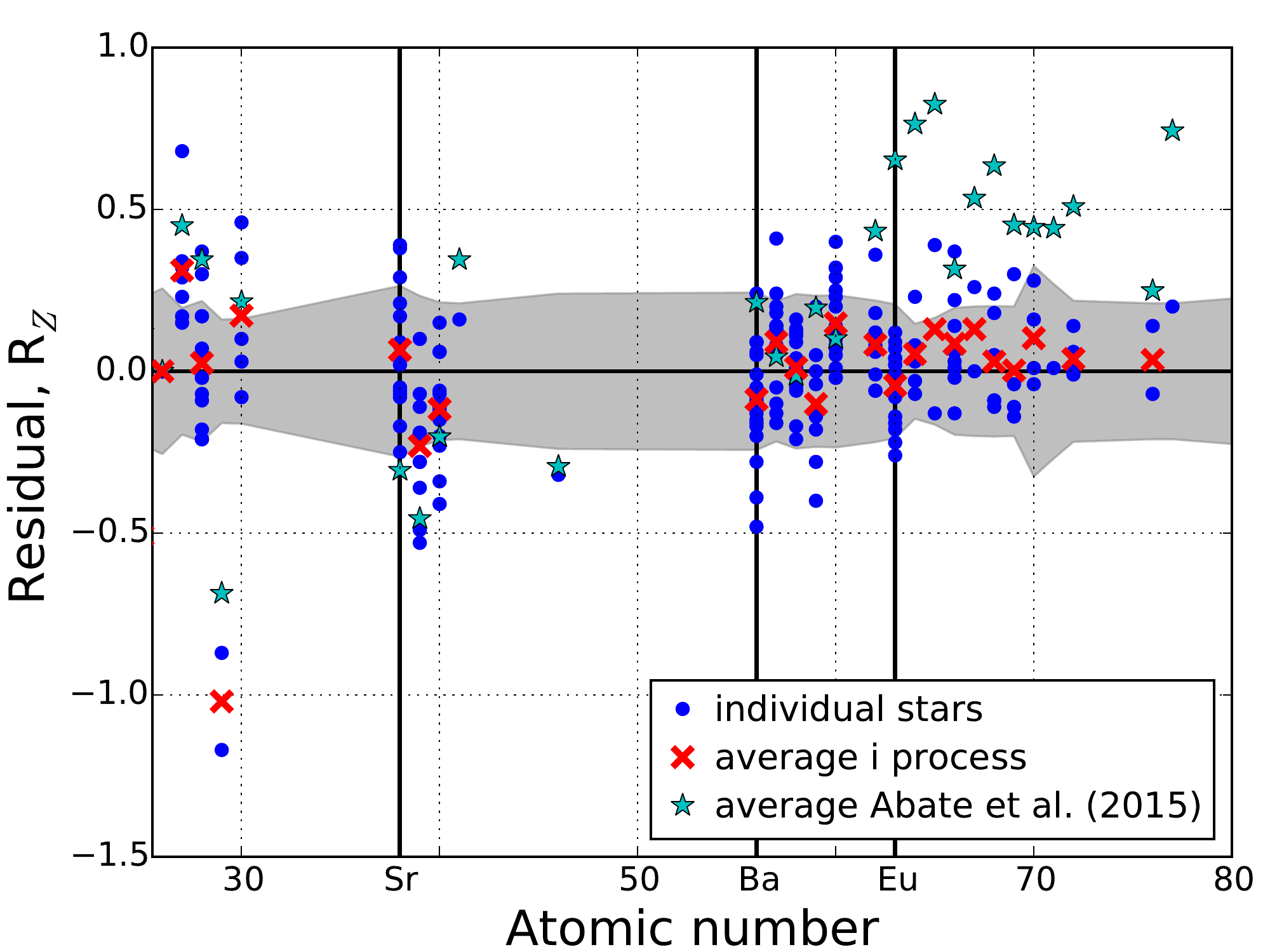}
\caption{Residuals in the best-fitting models for the surface abundances of CEMP-$s$/$r$ stars as defined in Eq. \ref{eq:residual}. The blue filled circles represent the residuals for individual stars and the average residual for each element is shown as red cross. The grey shaded area around $R_Z = 0$ indicates the uncertainty given by the errors in the abundance measurements averaged for each element. For comparison, the cyan stars show the average residuals of the best fits from \citet{Abate2015b}. }
\label{fig:cp_carlo_residuals}
\end{figure}

Fig. \ref{fig:cp_carlo_residuals} shows the residuals of every observed element for each star and their average.  
As a comparison, the average value of the residuals from \citet{Abate2015b} is shown as well.
The majority of the averaged residuals from this work lie within the observational uncertainty of the measurements. This result was not achieved previously when 
standard AGB nucleosynthesis only was used to model the surface abundances of CEMP-$s$/$r$ stars \citep{Abate2015b}.

For the heavy elements considered in the fits, only the average residual for Y lies on the 
boundary of the average measurement uncertainty, while the single Pd measurement is overpredicted 
in the corresponding best-fitting model to an extent only slightly outside the observational error. The remaining 
observed abundances of 21 elements with $30 < Z \leq 80$ are reproduced by the models of neutron capture 
nucleosynthesis with neutron densities in the $i$-process regime within the accuracy limited by the 
average errors in the measurements. We therefore believe the $i$ process is a valid component of the 
formation scenario of CEMP-$s$/$r$ stars. In particular, the possible connection of $i$-process 
nucleosynthesis and PIEs in low-metallicity AGB stars suggest a good candidate for a formation scenario of 
CEMP-$s$/$r$ stars in a binary system, analogous to the formation of CEMP-$s$ stars. This is supported by radial velocity measurements of CEMP-$s$/$r$ stars \citep{Hansen2016, Lucatello2009}. However, in this 
formation scenario it is likely that an AGB star with PIEs and $i$-process nucleosynthesis also undergoes 
``normal" thermal pulses with $s$-process nucleosynthesis. This means that the resulting heavy-element 
abundance pattern is most likely a superposition of an $s$- and $i$-process abundance pattern, instead of 
the pure $i$-process pattern studied in this work. An example of this might be the case of HE2148-1247 
where the $i$ process underestimates the Gd ($Z=64$) and Dy ($Z=66$) abundances (Fig.~\ref{fig:HE2148-1247}). In this case, dilution of the $i$-processed matter with $s$-processed matter will add additional Gd and Dy for the same total Ba (compare Fig.~\ref{fig:i-process-summary-XoverFe}).

\section{Summary, conclusions, and future work}

We have studied neutron-capture nucleosynthesis in an AGB-intershell region under the influence of constant 
neutron densities ranging from $s$-process conditions with $n = 10^{7}$\,cm$^{-3}$ up to $i$-process 
conditions with $n = 10^{15}$\,cm$^{-3}$. At high neutron densities, the balance between neutron-capture 
rates and $\beta$-decay rates for unstable isotopes changes in a way such that the neutron-capture path can 
flow further away from the valley of stability. This has two main consequences: (i) the bottleneck isotopes 
at magic neutron numbers are reached at lower atomic numbers compared to the $s$ process and (ii) the 
majority of isotopes that are produced along the neutron-capture path are unstable. For the $i$-process 
abundance pattern at $n = 10^{15}$\,cm$^{-3}$ the former means that a large amount of unstable \el{135}{I} 
with magic neutron number $N=50$ is produced while the latter means that all the \el{135}{I} decays into 
\el{135}{Ba} after the neutron source is switched off. While the equilibrium abundances of the ls-peak 
elements relative to iron are almost independent of the neutron density, the production of the hs-peak 
elements like Ba and $r$-process elements like Eu relative to iron increase with $n$ and the relative 
abundances internally within the ls and hs peaks are also strongly modified.

Comparing the results from the models to the surface abundances of a sample of 
20 CEMP-$s$/$r$ stars shows that the observed heavy element abundance patterns of all stars 
but one can be convincingly reproduced. Because the $i$-process models fit the abundances of 
CEMP-$s$/$r$ stars so well, we propose that this class should be renamed as CEMP-$i$ stars.
The majority of the best-fitting models have an 
abundance pattern created under the influence of a constant neutron density of $n = 
10^{14}$\,cm$^{-3}$. 
We stress that this work uses an extremely simplistic model.
The study of the influence of variations of the total neutron exposure on 
$i$-process abundance patterns goes beyond the scope of this work. In the future, 
the total neutron exposure should also be treated as a free parameter 
\citep[as done for example by][]{Roederer2016}. Including constraints from the observed 
lead abundances of CEMP-$s$/$r$ stars can then be used to explore the effects of the total neutron exposure on the $i$ process
and reveal more information about its potential physical sites. 
Additionally, deviations from a constant neutron-density profile need to be considered in future work, as well as additional thermal pulses with s-process nucleosynthesis. Moreover, a more realistic treatment (such as attempted by \citealt{Abate2015}) of mass transfer and dilution in the envelopes of both the AGB star and the companion is desirable.

Uncertainties in the rates of reactions important for the $i$ process were not considered 
in this study. \citet{Bertolli2013} studied how propagating systematic uncertainties of 
nuclear cross sections from different theoretical models changes the predicted abundance 
ratios of the hs elements and europium under $i$-process conditions. Depending on the 
theoretical model, changes of up to 1\,dex for [Ba/La] and $~0.5$\,dex for [La/Eu] are 
found \citep[see e.g. Fig. 6 of][]{Bertolli2013}. Therefore it is important to 
further study the influence of nuclear physics uncertainties and their influences on the 
predicted abundance patterns.

Finally, in the Large and Small Magellanic Clouds, post-AGB stars of low initial mass 
(1--1.5$M_{\odot}$) and metallicity $\left[ \mathrm{Fe}/\mathrm{H}\right] \approx -1$, 
higher than that of CEMP-$r$/$s$ stars, have recently been demonstrated to show abundance patterns 
incompatible with the $s$ process \citep{Lugaro2015}. These stars should be investigated 
in the light of the $i$ process to determine if this process is responsible for their 
peculiar abundances and to derive constraints on its metallicity dependence.

\acknowledgments
The authors thank Carlo Abate and Carolyn Doherty for very helpful discussion and the referee for her/his useful remarks. RJS is the recipient of a Sofja Kovalevskaja Award from the
Alexander von Humboldt Foundation. ML is a Momentum (``Lend\"ulet-2014'' Programme) 
project leader of the
Hungarian Academy of Sciences.

\bibliographystyle{apj}
\bibliography{references_with_data}

\newpage
\appendix

\section{$\alpha$ decay rates}
\begin{table}[htbp]
  \centering
   \caption{In the network included nuclear decay data from \citet{Tuli2011}. }
    \begin{tabular}{cccc}
    \toprule
    Decay & $t_{1/2} $ (s) & Decay Ratio (\%) & Decay Rate (s$^{-1}$) \\
    \tableline
    $^{5}\text{He} \to \text{n} + ^{4}\!\!\text{He}$  & $6.00\times10^{-25}$ & 100   & $1.16\times10^{24}$ \\
    $^{144}\text{Nd} \to ^{140}\!\!\text{Ce} + ^{4}\!\!\text{He}$  & $7.22\times10^{22}$ & 100   & $9.60\times10^{-24}$ \\
    $^{145}\text{Pm} \to ^{141}\!\!\text{Pr} + ^{4}\!\!\text{He}$  & $5.58\times10^{8}$ & $2.80\times10^{-7}$ & $3.48\times10^{-18}$ \\
    $^{146}\text{Sm} \to ^{142}\!\!\text{Nd} + ^{4}\!\!\text{He}$  & $3.25\times10^{15}$ & 100   & $2.13\times10^{-16}$ \\
    $^{147}\text{Sm} \to ^{143}\!\!\text{Nd} + ^{4}\!\!\text{He}$  & $3.34\times10^{18}$ & 100   & $2.07\times10^{-19}$ \\
    $^{148}\text{Sm} \to ^{144}\!\!\text{Nd} + ^{4}\!\!\text{He}$  & $2.21\times10^{23}$ & 100   & $3.14\times10^{-24}$ \\
    $^{150}\text{Gd} \to ^{146}\!\!\text{Sm} + ^{4}\!\!\text{He}$  & $5.64\times10^{13}$ & 100   & $1.23\times10^{-14}$ \\
    $^{151}\text{Gd} \to ^{147}\!\!\text{Sm} + ^{4}\!\!\text{He}$  & $1.07\times10^{7}$ & $8.00\times10^{-7}$ & $5.18\times10^{-16}$ \\
    $^{152}\text{Gd} \to ^{148}\!\!\text{Sm} + ^{4}\!\!\text{He}$  & $3.41\times10^{21}$ & 100   & $2.04\times10^{-22}$ \\
    $^{152}\text{Dy} \to ^{148}\!\!\text{Gd} + ^{4}\!\!\text{He}$  & $8.57\times10^{3}$ & 0.1   & $8.09\times10^{-8}$ \\
    $^{153}\text{Dy} \to ^{149}\!\!\text{Gd} + ^{4}\!\!\text{He}$  & $2.30\times10^{4}$ & $9.40\times10^{-3}$ & $2.83\times10^{-9}$ \\
    $^{154}\text{Dy} \to ^{150}\!\!\text{Gd} + ^{4}\!\!\text{He}$  & $9.46\times10^{13}$ & 100   & $7.33\times10^{-15}$ \\
    $^{187}\text{Re} \to ^{183}\!\!\text{Ta} + ^{4}\!\!\text{He}$  & $1.37\times10^{18}$ & $1.00\times10^{-4}$ & $5.08\times10^{-25}$ \\
    $^{210}\text{Pb} \to ^{206}\!\!\text{Hg} + ^{4}\!\!\text{He}$  & $7.00\times10^{8}$ & $1.90\times10^{-6}$ & $1.88\times10^{-17}$ \\
    $^{210}\text{Bi} \to ^{206}\!\!\text{Tl} + ^{4}\!\!\text{He}$  & $4.33\times10^{5}$ & $1.30\times10^{-4}$ & $2.08\times10^{-12}$ \\
    $^{211}\text{Bi} \to ^{207}\!\!\text{Tl} + ^{4}\!\!\text{He}$  & $1.28\times10^{2}$ & 99.72 & $5.38\times10^{-3}$ \\
    $^{212}\text{Bi} \to ^{208}\!\!\text{Tl} + ^{4}\!\!\text{He}$  & $3.63\times10^{3}$ & 35.94 & $6.86\times10^{-5}$ \\
    $^{213}\text{Bi} \to ^{209}\!\!\text{Tl} + ^{4}\!\!\text{He}$  & $2.74\times10^{3}$ & 2.2   & $5.57\times10^{-6}$ \\
    $^{214}\text{Bi} \to ^{210}\!\!\text{Tl} + ^{4}\!\!\text{He}$  & $1.19\times10^{3}$ & 0.02  & $1.16\times10^{-7}$ \\
    $^{210}\text{Po} \to ^{206}\!\!\text{Pb} + ^{4}\!\!\text{He}$  & $1.20\times10^{7}$ & 100   & $5.80\times10^{-8}$ \\
    $^{211}\text{Po} \to ^{207}\!\!\text{Pb} + ^{4}\!\!\text{He}$  & $5.16\times10^{-1}$ & 100   & 1.34 \\
    $^{212}\text{Po} \to ^{208}\!\!\text{Pb} + ^{4}\!\!\text{He}$  & $2.99\times10^{-7}$ & 100   & $2.32\times10^{6}$ \\
    $^{213}\text{Po} \to ^{209}\!\!\text{Pb} + ^{4}\!\!\text{He}$  & $3.72\times10^{-6}$ & 100   & $1.86\times10^{5}$ \\
    $^{214}\text{Po} \to ^{210}\!\!\text{Pb} + ^{4}\!\!\text{He}$  & $1.64\times10^{-4}$ & 100   & $4.22\times10^{3}$ \\
    $^{215}\text{Po} \to ^{211}\!\!\text{Pb} + ^{4}\!\!\text{He}$  & $1.78\times10^{-3}$ & 100   & $3.89\times10^{2}$ \\
    $^{216}\text{Po} \to ^{212}\!\!\text{Pb} + ^{4}\!\!\text{He}$  & $1.45\times10^{-1}$ & 100   & 4.78 \\
    $^{217}\text{Po} \to ^{213}\!\!\text{Pb} + ^{4}\!\!\text{He}$  & 1.53 & 100   & $4.53\times10^{-1}$ \\
    $^{218}\text{Po} \to ^{214}\!\!\text{Pb} + ^{4}\!\!\text{He}$  & $1.86\times10^{2}$ & 99.98 & $3.73\times10^{-3}$ \\
    \tableline    
    \end{tabular}%
  \label{tab:alphadecays}%
\end{table}%

\newpage
\clearpage
\section{Tables}

\begin{table}[htbp]
  \centering
  \caption{Final  [X/Fe] for the simulations with different neutron densities $n$.}
    \begin{tabular}{cccccc}
    \toprule
    &  & \multicolumn{4}{c}{Neutron density in cm$^{-3}$} \\
Z & element & $10^{12}$& $10^{13}$& $10^{14}$ & $10^{15}$  \\
\tableline
26 & Fe & 0.00 & 0.00 & 0.00 & 0.00 \\
27 & Co & 1.76 & 1.92 & 2.02 & 1.90 \\
28 & Ni & 1.43 & 1.60 & 1.71 & 1.60 \\
29 & Cu & 2.55 & 2.75 & 2.89 & 2.80 \\
30 & Zn & 2.36 & 2.86 & 3.22 & 3.13 \\
31 & Ga & 3.08 & 3.36 & 3.58 & 3.71 \\
32 & Ge & 3.16 & 3.38 & 3.61 & 3.73 \\
33 & As & 3.21 & 3.73 & 4.06 & 4.33 \\
34 & Se & 3.40 & 3.77 & 4.28 & 4.29 \\
35 & Br & 3.33 & 3.64 & 4.01 & 4.09 \\
36 & Kr & 4.79 & 4.94 & 5.04 & 4.90 \\
37 & Rb & 4.66 & 4.79 & 4.84 & 4.81 \\
38 & Sr & 4.31 & 4.32 & 4.35 & 4.37 \\
39 & Y & 4.52 & 4.65 & 4.58 & 4.64 \\
40 & Zr & 4.74 & 4.90 & 5.02 & 4.88 \\
41 & Nb & 4.54 & 4.64 & 5.05 & 5.25 \\
42 & Mo & 4.80 & 5.16 & 5.33 & 5.25 \\
44 & Ru & 4.51 & 4.72 & 4.96 & 5.04 \\
45 & Rh & 4.32 & 4.47 & 4.67 & 4.82 \\
46 & Pd & 4.80 & 5.03 & 5.16 & 5.02 \\
47 & Ag & 4.22 & 4.48 & 4.70 & 4.85 \\
48 & Cd & 4.78 & 5.05 & 5.19 & 5.06 \\
49 & In & 4.86 & 5.03 & 5.13 & 5.10 \\
50 & Sn & 5.11 & 5.30 & 5.40 & 5.25 \\
51 & Sb & 5.08 & 5.26 & 5.36 & 5.24 \\
52 & Te & 4.96 & 5.24 & 5.40 & 5.40 \\
53 & I & 4.65 & 5.18 & 5.47 & 5.44 \\
54 & Xe & 5.36 & 5.87 & 6.06 & 6.14 \\
55 & Cs & 4.98 & 5.37 & 6.04 & 6.53 \\
56 & Ba & 5.42 & 5.55 & 6.06 & 6.57 \\
57 & La & 5.38 & 5.53 & 5.66 & 5.59 \\
58 & Ce & 5.12 & 5.36 & 5.76 & 5.72 \\
59 & Pr & 5.30 & 5.40 & 5.81 & 5.87 \\
60 & Nd & 5.39 & 5.54 & 5.63 & 5.57 \\
62 & Sm & 5.10 & 5.30 & 5.65 & 5.75 \\
63 & Eu & 4.79 & 5.16 & 5.48 & 5.56 \\
64 & Gd & 5.05 & 5.36 & 5.63 & 5.63 \\
65 & Tb & 4.85 & 5.00 & 5.13 & 5.35 \\
66 & Dy & 4.88 & 5.01 & 5.31 & 5.41 \\
67 & Ho & 4.76 & 5.02 & 5.18 & 5.22 \\
68 & Er & 5.29 & 5.54 & 5.72 & 5.78 \\
69 & Tm & 5.14 & 5.32 & 5.60 & 5.80 \\
70 & Yb & 5.61 & 5.80 & 5.85 & 5.80 \\
71 & Lu & 5.24 & 5.57 & 5.99 & 5.90 \\
72 & Hf & 5.54 & 5.80 & 6.04 & 6.00 \\
73 & Ta & 5.88 & 6.03 & 6.06 & 6.12 \\
74 & W & 5.59 & 5.90 & 6.10 & 6.16 \\
75 & Re & 5.46 & 5.69 & 5.88 & 5.90 \\
76 & Os & 5.14 & 5.31 & 5.37 & 5.35 \\
77 & Ir & 4.35 & 4.56 & 4.83 & 4.91 \\
78 & Pt & 4.95 & 5.17 & 5.37 & 5.33 \\
79 & Au & 4.78 & 4.97 & 5.20 & 5.19 \\
80 & Hg & 5.73 & 6.02 & 6.32 & 6.28 \\
\tableline
    \end{tabular}%
  \label{tab:final_abu}%
\end{table}%

\begin{table}[htbp]
  \centering
  \caption{Details about the 20 CEMP-$s$/$r$ stars in the sample of \citet{Abate2015b}: surface gravities, temperatures and selected chemical properties.}
    \begin{tabular}{ccccccccc}
    \toprule
ID & $\log \left( g/\text{cm\,s}^{-2} \right) $  & $T_{\text{eff}}$  (K) & Number of & [Fe/H] & [C/Fe] & [Ba/Fe] & [Eu/Fe] & Reference \\
 & & & observed elements & & & & & \\
\tableline
BS16080-175 & 3.7(2) & 6240 & 6 & -1.9 & 1.8 & 1.6 & 1.1 & 1 \\
BS17436-058 & 2.7(2) & 5690 & 7 & -1.8 & 1.6 & 1.7 & 1.2 & 1 \\
CS22881-036 & 4.0(1) & 6200 & 14 & -2.1 & 2.1 & 1.9 & 1.0 & 23 \\
CS22898-027 & 3.7(3) & 6110 & 22 & -2.3 & 2.0 & 2.3 & 2.0 & 3, 4, 18, 20 \\
CS22948-027 & 1.8(4) & 4800 & 21 & -2.5 & 2.4 & 2.4 & 1.9 & 5, 7 \\
CS29497-030 & 4.0(5) & 6966 & 33 & -2.5 & 2.4 & 2.3 & 1.7 & 15, 16, 24 \\
CS29526-110 & 3.2(1) & 6500 & 18 & -2.4 & 2.3 & 2.1 & 1.8 & 3, 4 \\
CS31062-012 & 4.2(4) & 6099 & 24 & -2.8 & 2.3 & 2.1 & 1.6 & 2, 3, 4, 6, 21 \\
CS31062-050 & 2.9(3) & 5489 & 37 & -2.5 & 1.9 & 2.4 & 2.0 & 2, 3, 4, 6 \\
HD187861 & 2.0(4) & 4960 & 14 & -2.4 & 2.0 & 1.9 & 1.3 & 19, 26 \\
HD224959 & 1.9(3) & 5050 & 14 & -2.1 & 1.8 & 2.2 & 1.7 & 19 \\
HE0131-3953 & 3.8(1) & 5928 & 16 & -2.7 & 2.5 & 2.2 & 1.7 & 8 \\
HE0143-0441 & 4.0(4) & 6305 & 22 & -2.4 & 2.0 & 2.4 & 1.7 & 12, 13 \\
HE0338-3945 & 4.1(4) & 6161 & 32 & -2.5 & 2.1 & 2.4 & 2.0 & 8, 17 \\
HE1105+0027 & 3.5(1) & 6132 & 16 & -2.5 & 2.0 & 2.4 & 1.9 & 8 \\
HE1305+0007 & 1.5(5) & 4655 & 21 & -2.2 & 2.1 & 2.6 & 2.2 & 9, 14 \\
HE2148-1247 & 3.9(1) & 6380 & 25 & -2.4 & 2.0 & 2.4 & 2.0 & 11 \\
HE2258-6358 & 1.6(1) & 4900 & 31 & -2.7 & 2.4 & 2.3 & 1.7 & 22 \\
LP625-44 & 2.6(3) & 5500 & 31 & -2.8 & 2.3 & 2.8 & 1.9 & 2, 4, 21 \\
SDSSJ0912+0216 & 4.5(1) & 6500 & 28 & -2.6 & 2.3 & 1.6 & 1.3 & 10 \\
\tableline
    \end{tabular}%
\tablerefs{(1) \citealt{Allen2012}; (2) \citealt{Aoki2001}; (3) \citealt{Aoki2002a}; (4) \citealt{Aoki2002b}; (5) \citealt{Aoki2007}; (6) \citealt{Aoki2008}; (7) \citealt{Barbuy2005}; (8) \citealt{Barklem2005}; (9) \citealt{Beers2007}; (10) \citealt{Behara2010}; (11) \citealt{Cohen2003}; (12) \citealt{Cohen2004}; (13) \citealt{Cohen2006}; (14) \citealt{Goswami2006}; (15) \citealt{Ivans2005}; (16) \citealt{Johnson2007}; (17) \citealt{Jonsell2006}; (18) \citealt{Lai2007}; (19) \citealt{Masseron2010}; (20) \citealt{McWilliam1995}; (21) \citealt{Norris1997}; (22) \citealt{Placco2013}; (23) \citealt{Preston2001}; (24) \citealt{Sivarani2004}; (25) \citealt{Sneden2003b}; (26) \citealt{vanEck2003}}
  \label{tab:data_sample}%
\end{table}%

\newpage
\clearpage
\section{All fits}
\label{sec:all_fits}

This section shows the best fitting models for each of the 20 CEMP-$s$/$r$ stars. Details of each 
best fit (neutron density $n$, $\chi^2$ and dilution factor $d$) are shown in the right corner of the 
plots. The lower panel shows the distribution of the residuals. The uncertainty of the observations 
$\sigma_{Z,\text{obs}}$ is indicated by errorbars in the upper panel and by the shaded region in the 
lower panel. The vertical lines show the location of Sr, Ba and Eu which are representatives of the 
ls and hs peak as well as the $r$ process, respectively.
We note that the abundance patterns reported by \citet{Behara2010} for SDSSJ0912+0216 (as well as 
SDSSJ1349-0229, which is not included in the sample of \citet{Abate2015b} because of the low 
[Fe/H]) are at odds with any $s$ or $i$-process models and need an independent verification.

\begin{figure}[h]
\begin{minipage}[t]{.46\textwidth}
\centering
\includegraphics[ width=\linewidth]{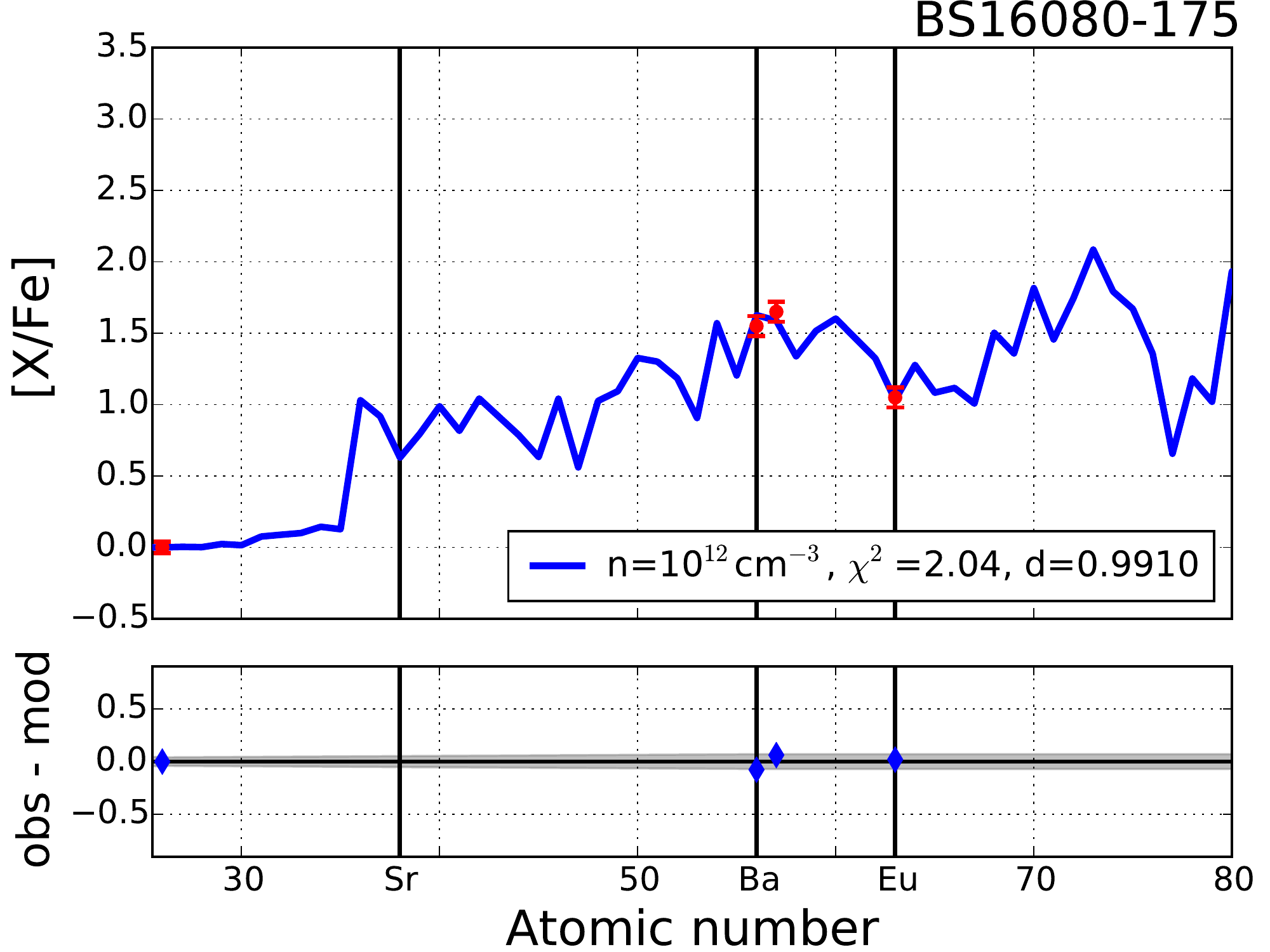}
\caption{Best fitting model for CEMP-$s$/$r$ star BS16080-175. The best fitting s-process models with initial $r$-process enhancement can be found in Fig.~5 from \citet{Bisterzo2012}.}
\end{minipage}
\hfill
\begin{minipage}[t]{.46\textwidth}
\centering
\includegraphics[ width=\linewidth]{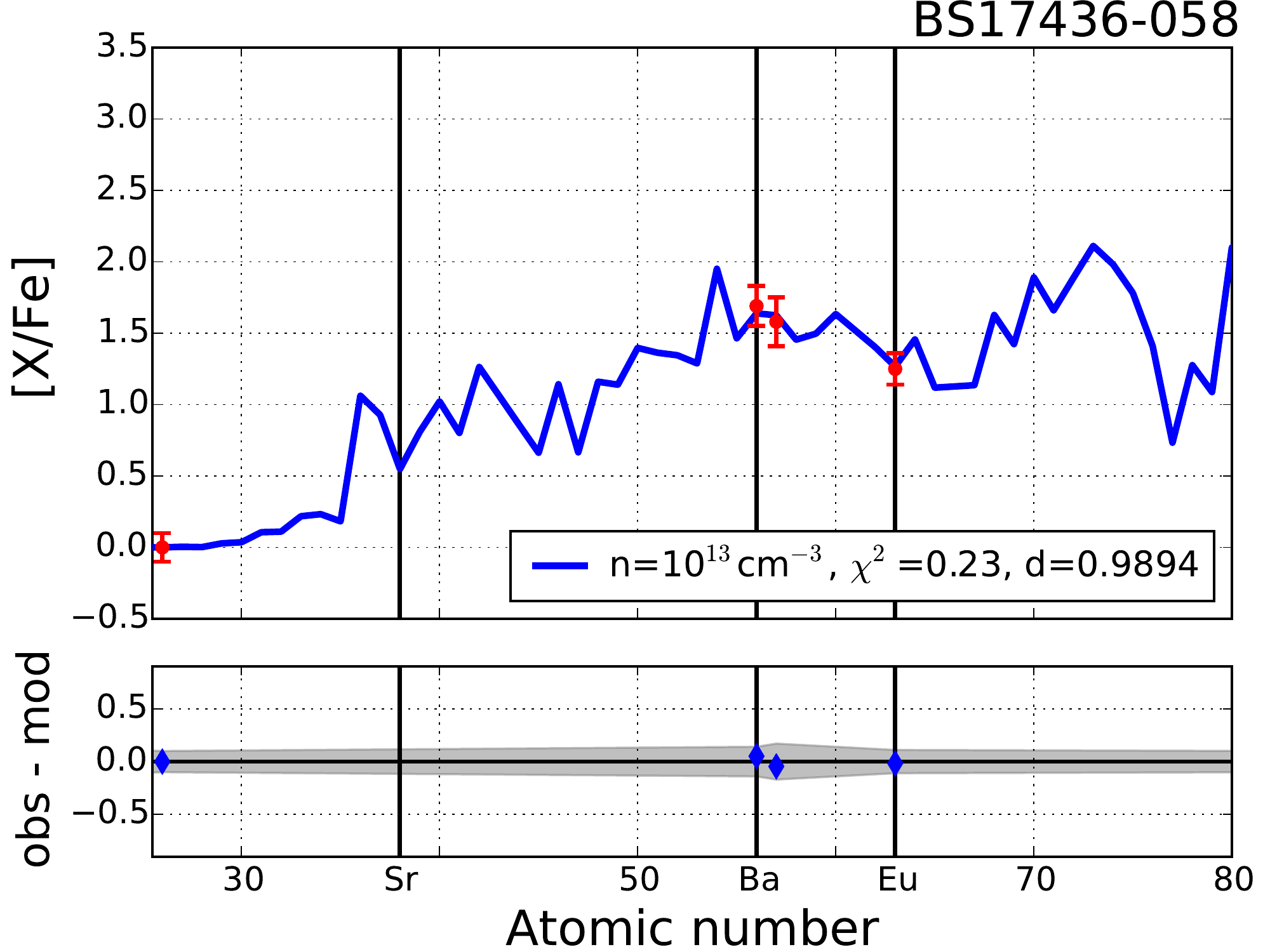}
\caption{Best fitting model for CEMP-$s$/$r$ star BS17436-058. The best fitting s-process models with initial $r$-process enhancement can be found in Fig.~16 from \citet{Bisterzo2012}.}
\end{minipage}
\hfill
\begin{minipage}[t]{.46\textwidth}
\vspace{12pt}
\centering
\includegraphics[ width=\linewidth]{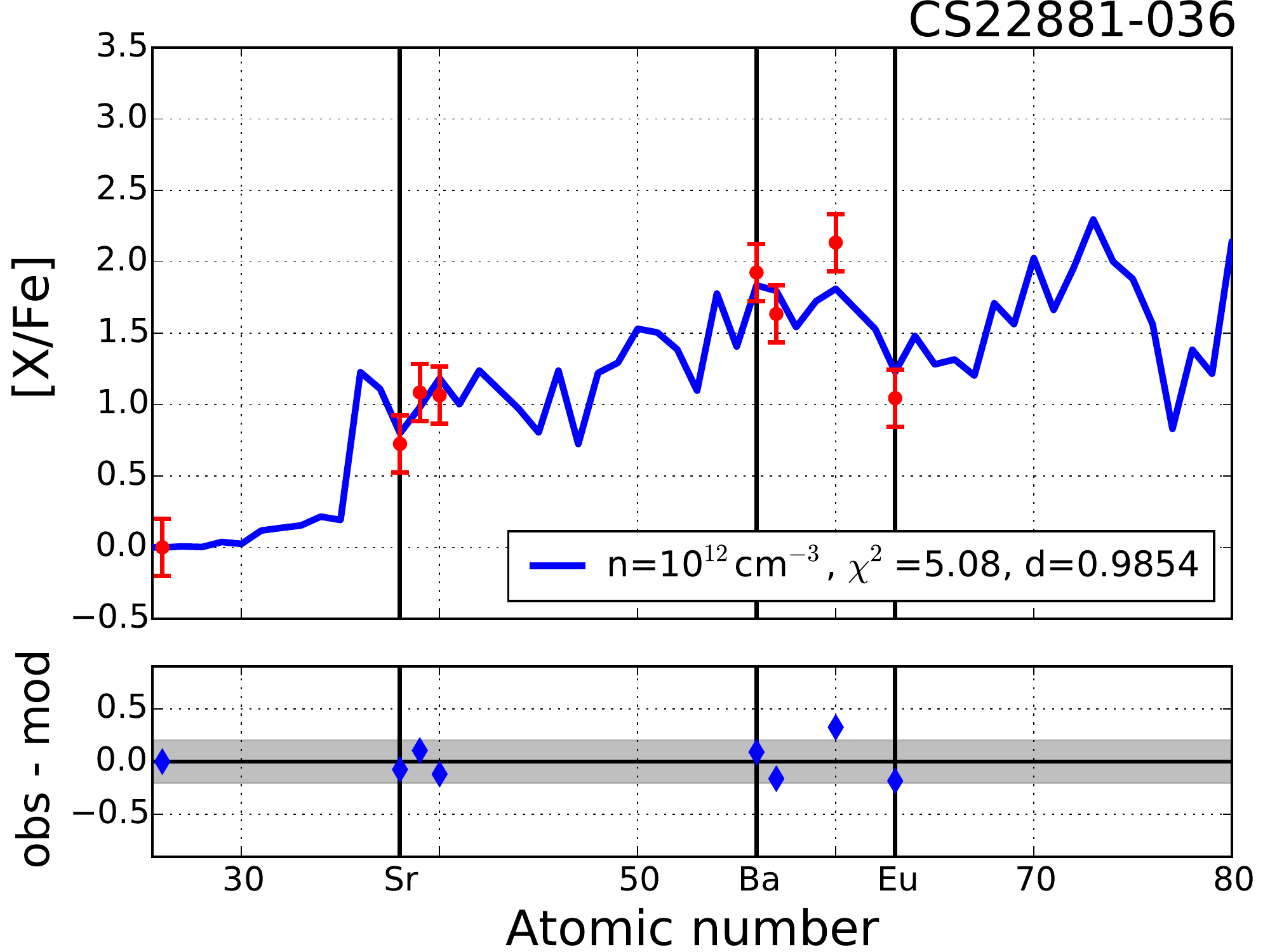}
\caption{Best fitting model for CEMP-$s$/$r$ star CS22881-036. The best fitting s-process models with initial $r$-process enhancement can be found in Fig.~1 from \citet{Bisterzo2012}.}
\end{minipage}
\hfill
\begin{minipage}[t]{.46\textwidth}
\vspace{12pt}
\centering
\includegraphics[ width=\linewidth]{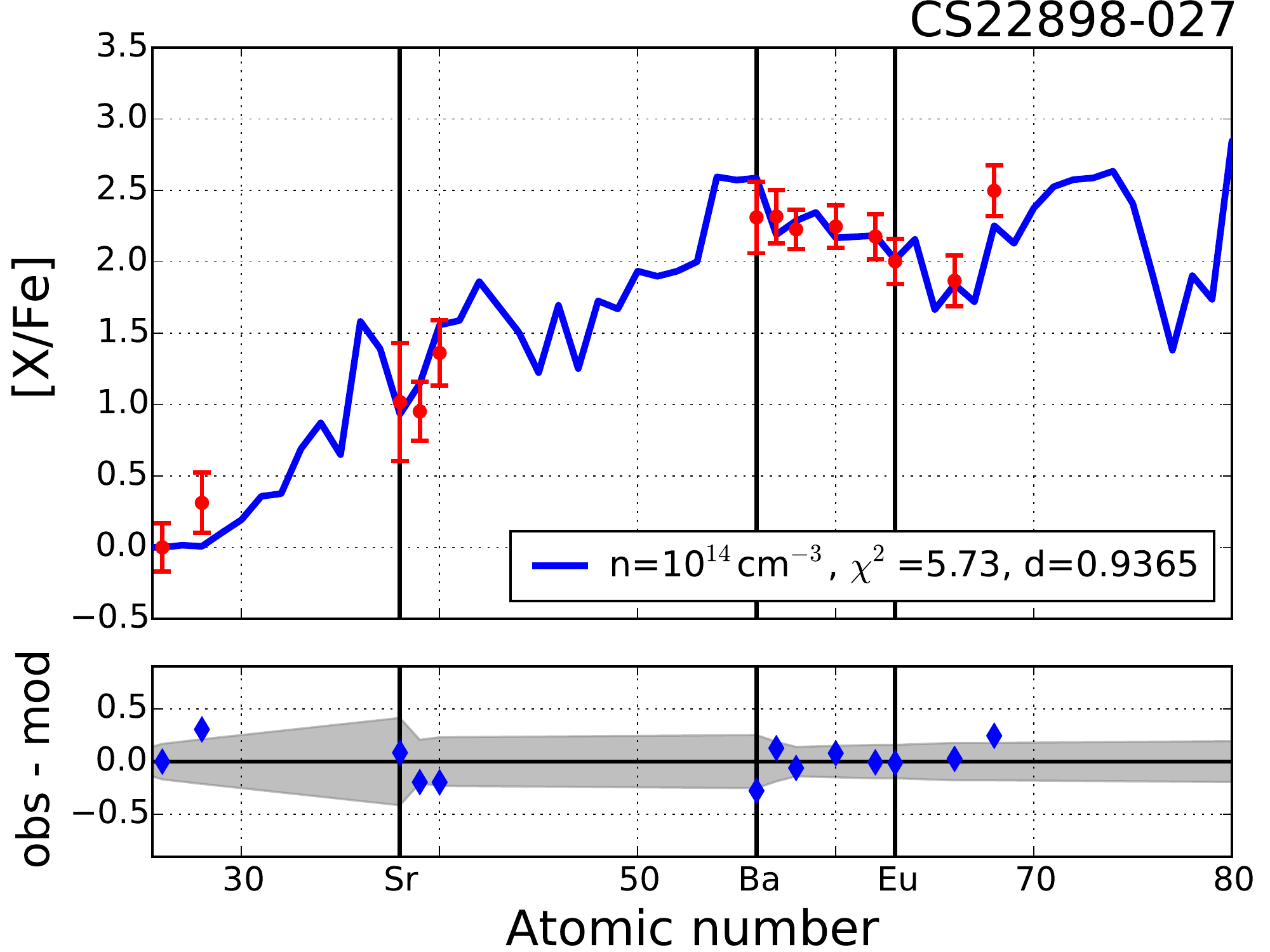}
\caption{Best fitting model for CEMP-$s$/$r$ star CS22898-027. The best fitting s-process models with initial $r$-process enhancement can be found in Fig.~17 from \citet{Bisterzo2012} and with binary evolution in Fig.~5 from \citet{Abate2015b}.}
\end{minipage}
\end{figure}
\begin{figure}
\begin{minipage}[t]{.46\textwidth}
\vspace{12pt}
\centering
\includegraphics[ width=\linewidth]{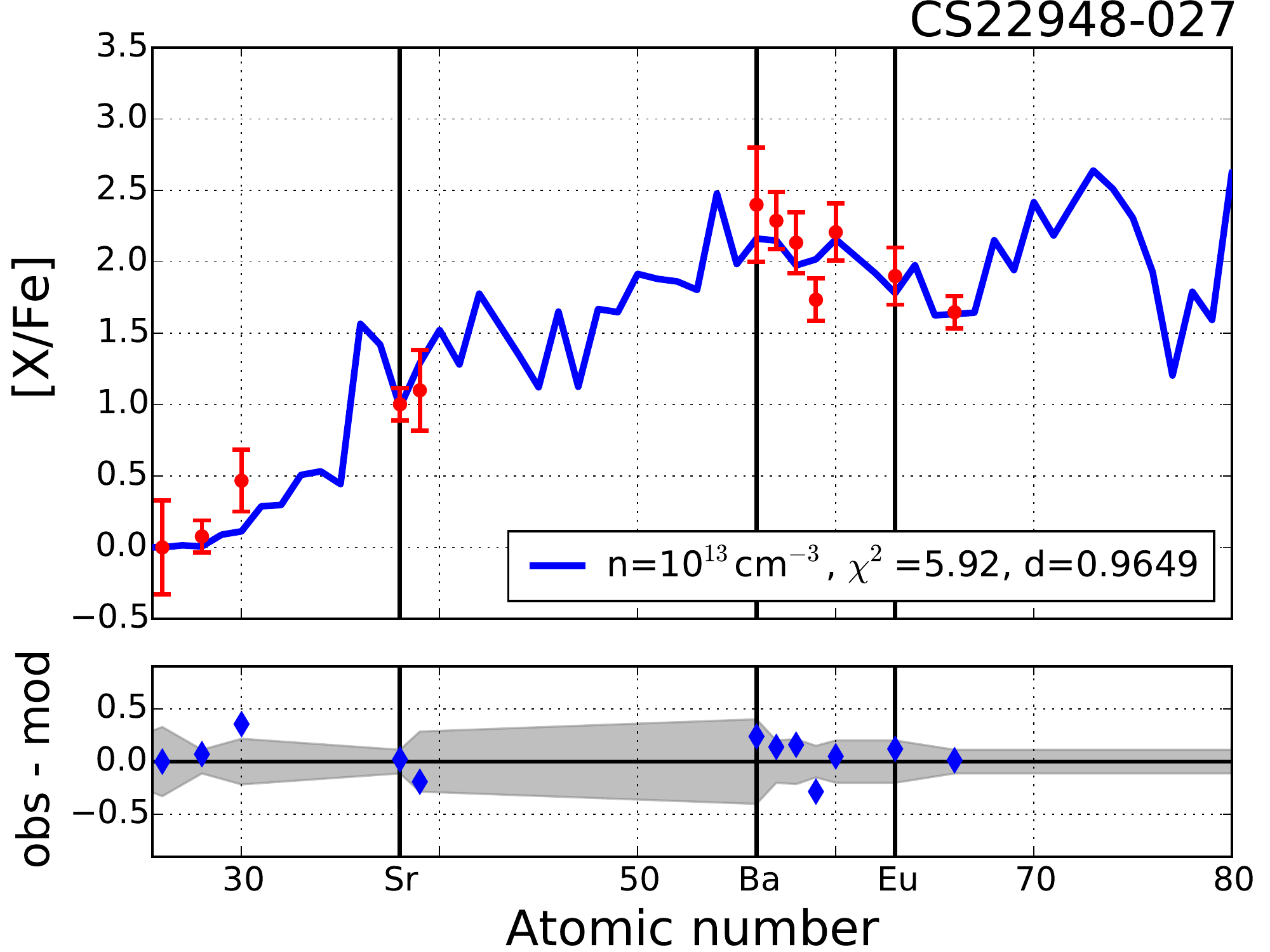}
\caption{Best fitting model for CEMP-$s$/$r$ star CS22948-027. The best fitting s-process models with initial $r$-process enhancement can be found in Fig.~27 from \citet{Bisterzo2012} and with binary evolution in Fig.~A2 from \citet{Abate2015}.}
\end{minipage}
\hfill
\begin{minipage}[t]{.46\textwidth}
\vspace{12pt}
\centering
\includegraphics[ width=\linewidth]{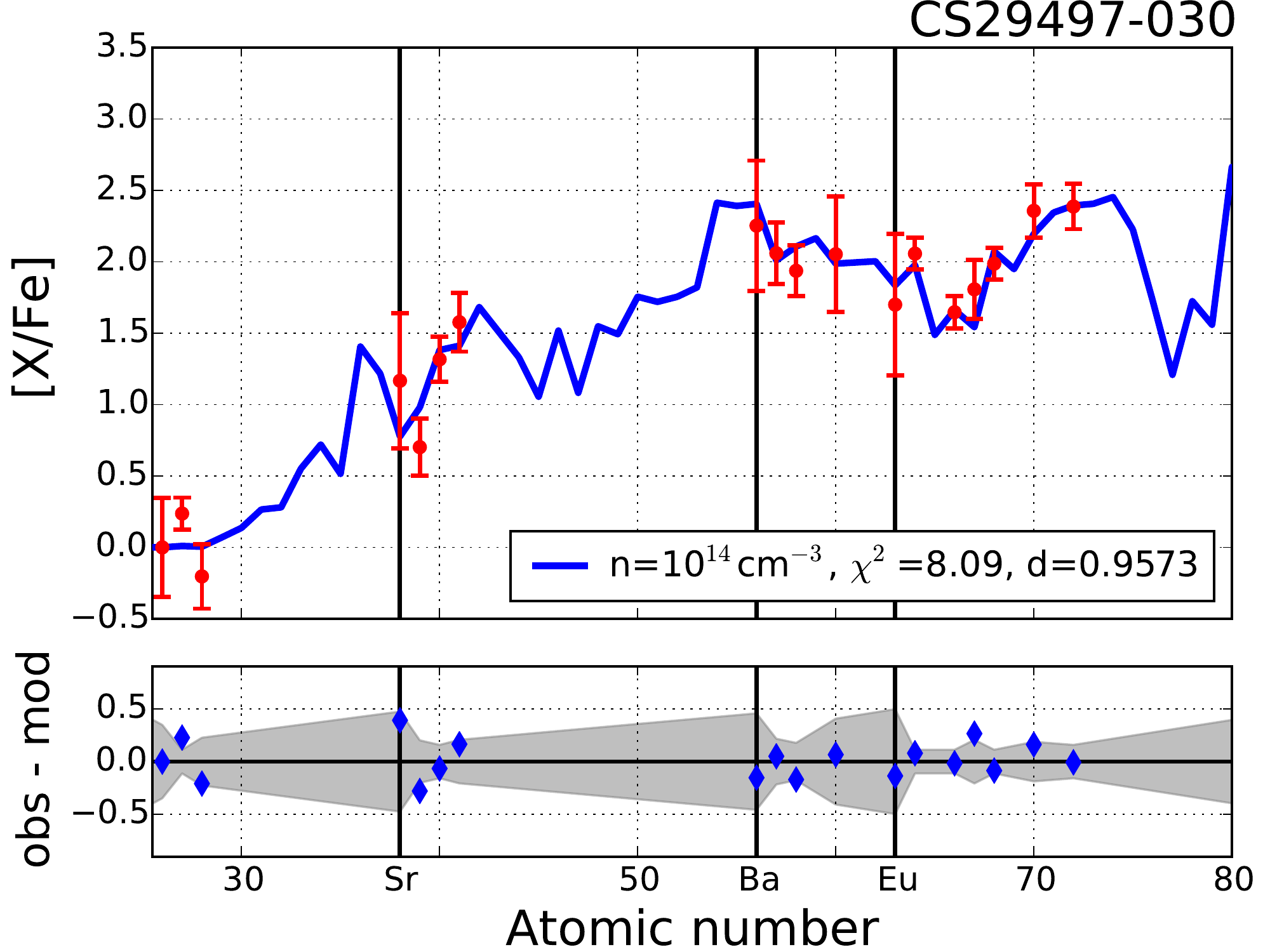}
\caption{Best fitting model for CEMP-$s$/$r$ star CS29497-030. The best fitting s-process models with initial $r$-process enhancement can be found in Fig.~18 from \citet{Bisterzo2012} and with binary evolution in Fig.~6 from \citet{Abate2015}.}
\end{minipage}
\hfill
\begin{minipage}[t]{.46\textwidth}
\vspace{12pt}
\centering
\includegraphics[ width=\linewidth]{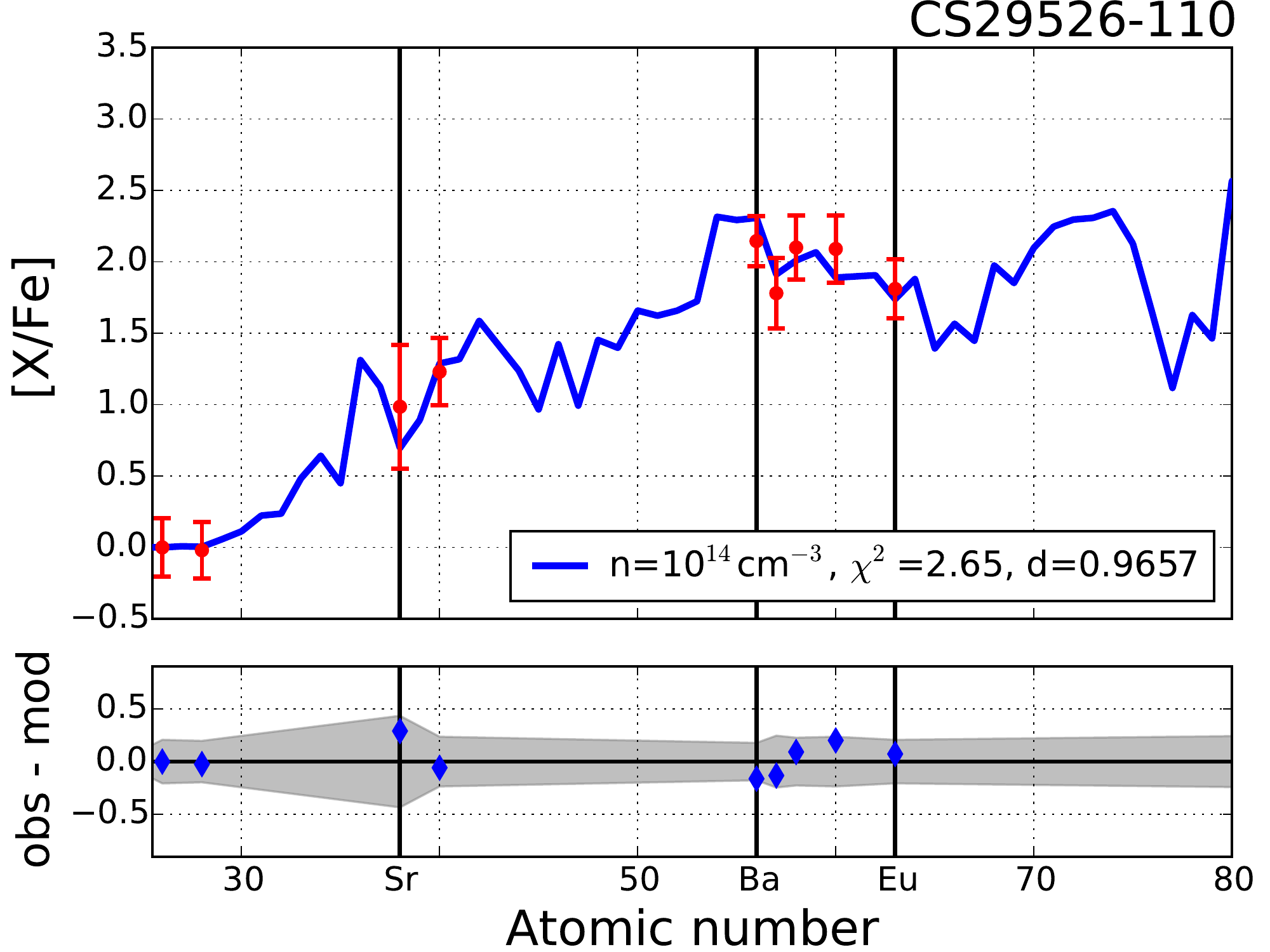}
\caption{Best fitting model for CEMP-$s$/$r$ star CS29526-110. The best fitting s-process models with initial $r$-process enhancement can be found in Fig.~23 from \citet{Bisterzo2012}.}
\end{minipage}
\hfill
\begin{minipage}[t]{.46\textwidth}
\vspace{12pt}
\centering
\includegraphics[ width=\linewidth]{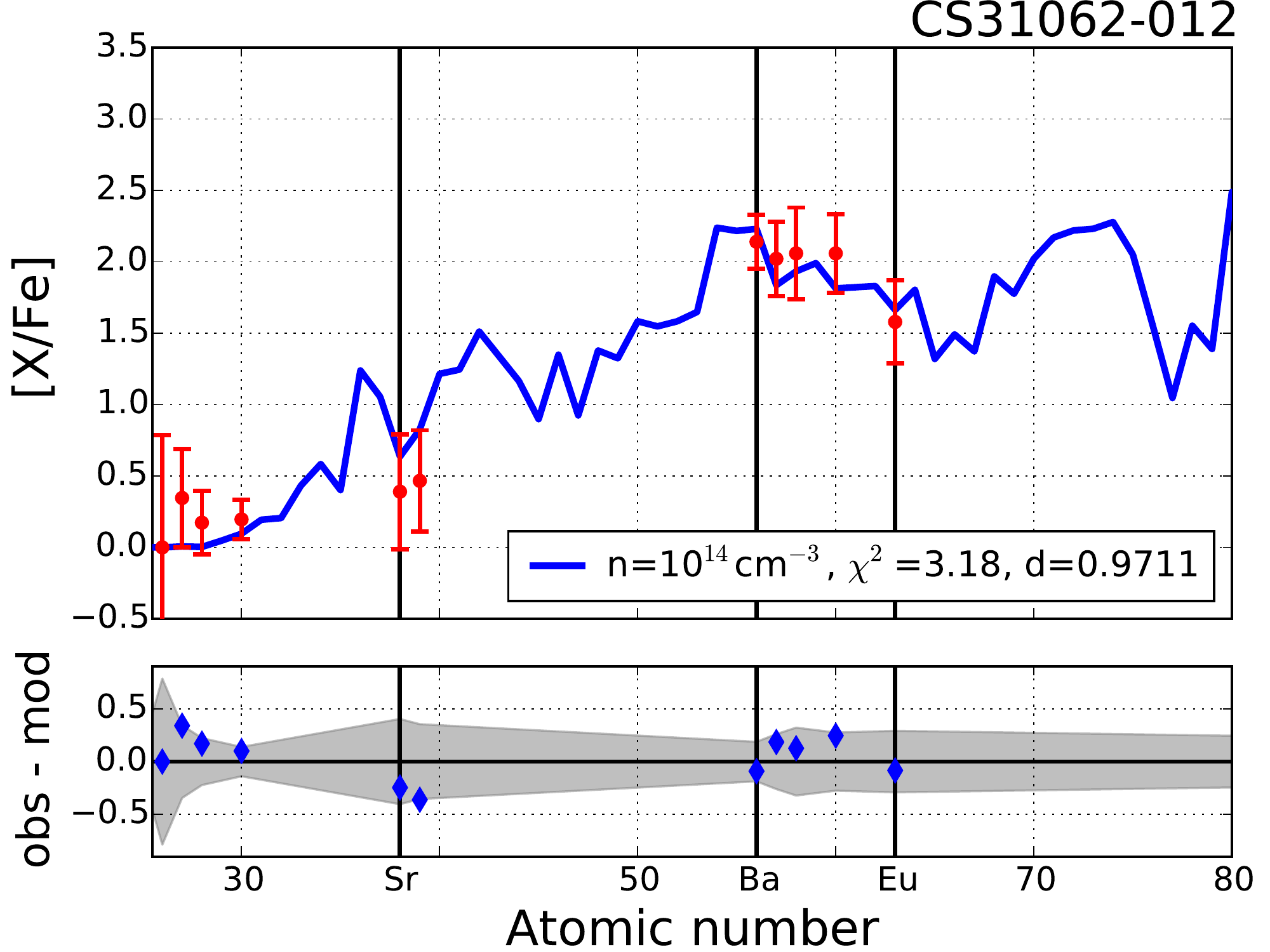}
\caption{Best fitting model for CEMP-$s$/$r$ star CS31062-012. The best fitting s-process models with initial $r$-process enhancement can be found in Fig.~24 from \citet{Bisterzo2012}.}
\end{minipage}
\hfill
\begin{minipage}[t]{.46\textwidth}
\vspace{12pt}
\centering
\includegraphics[ width=\linewidth]{CS31062-050-fit.pdf}
\caption{Best fitting model for CEMP-$s$/$r$ star CS31062-050. The best fitting s-process models with initial $r$-process enhancement can be found in Fig.~26 from \citet{Bisterzo2012}.}
\end{minipage}
\hfill
\begin{minipage}[t]{.46\textwidth}
\vspace{12pt}
\centering
\includegraphics[ width=\linewidth]{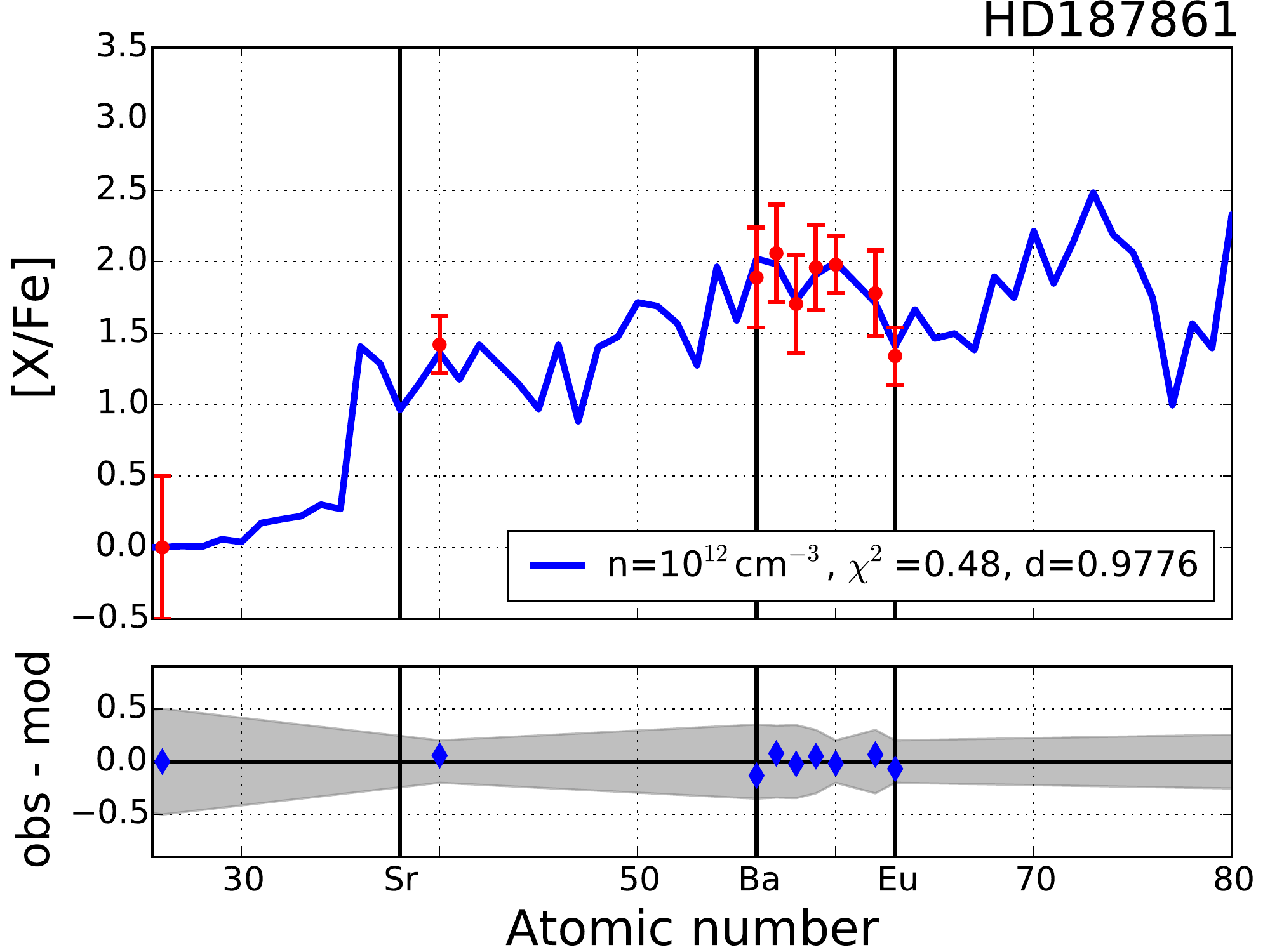}
\caption{Best fitting model for CEMP-$s$/$r$ star HD187861. The best fitting s-process models with initial $r$-process enhancement can be found in Fig.~29 from \citet{Bisterzo2012}.}
\end{minipage}
\end{figure}
\begin{figure}
\begin{minipage}[t]{.46\textwidth}
\vspace{12pt}
\centering
\includegraphics[ width=\linewidth]{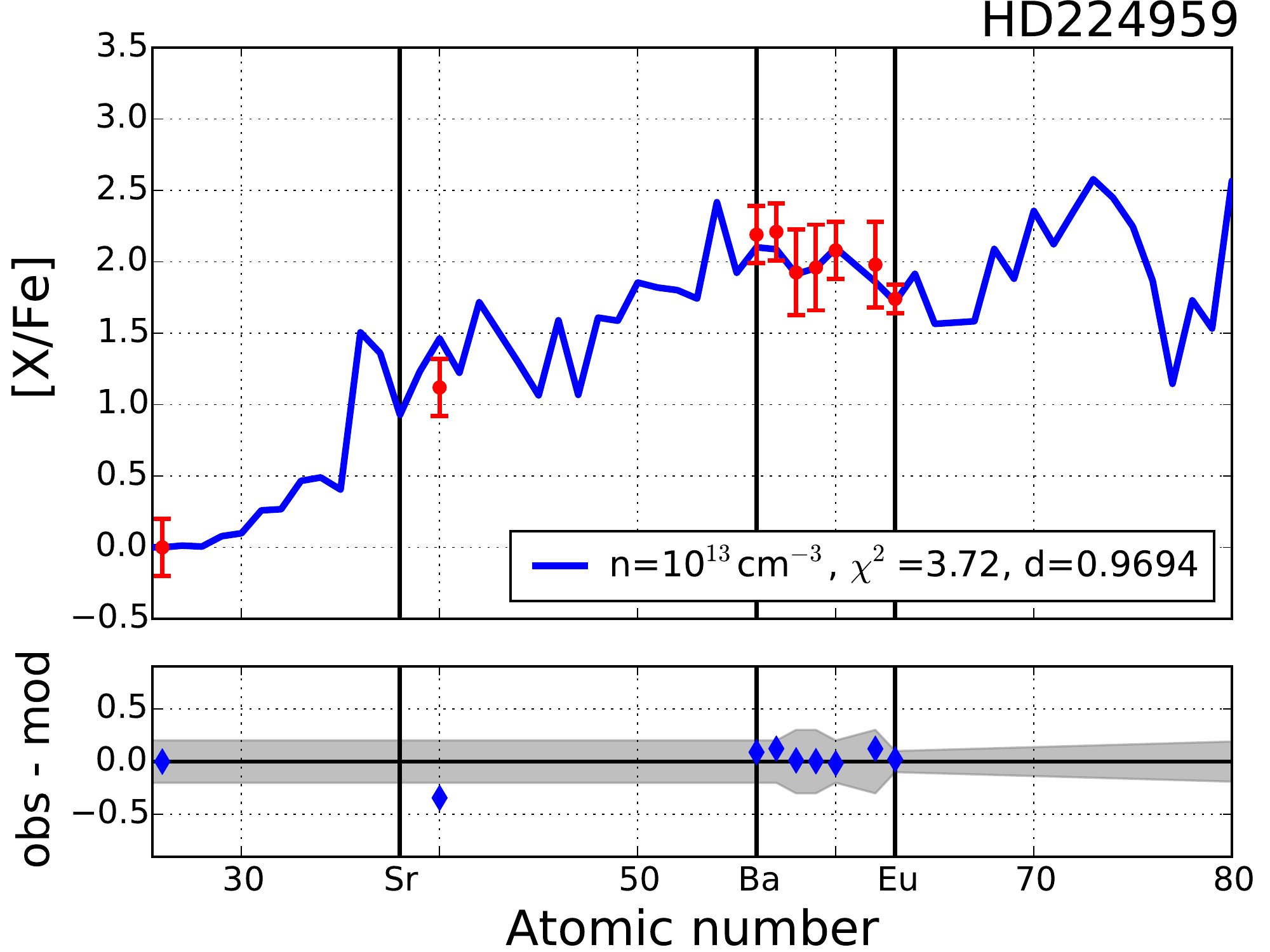}
\caption{Best fitting model for CEMP-$s$/$r$ star HD224959. The best fitting s-process models with initial $r$-process enhancement can be found in Fig.~30 from \citet{Bisterzo2012} and with binary evolution in Fig.~A8 from \citet{Abate2015}.}
\end{minipage}
\hfill
\begin{minipage}[t]{.46\textwidth}
\vspace{12pt}
\centering
\includegraphics[ width=\linewidth]{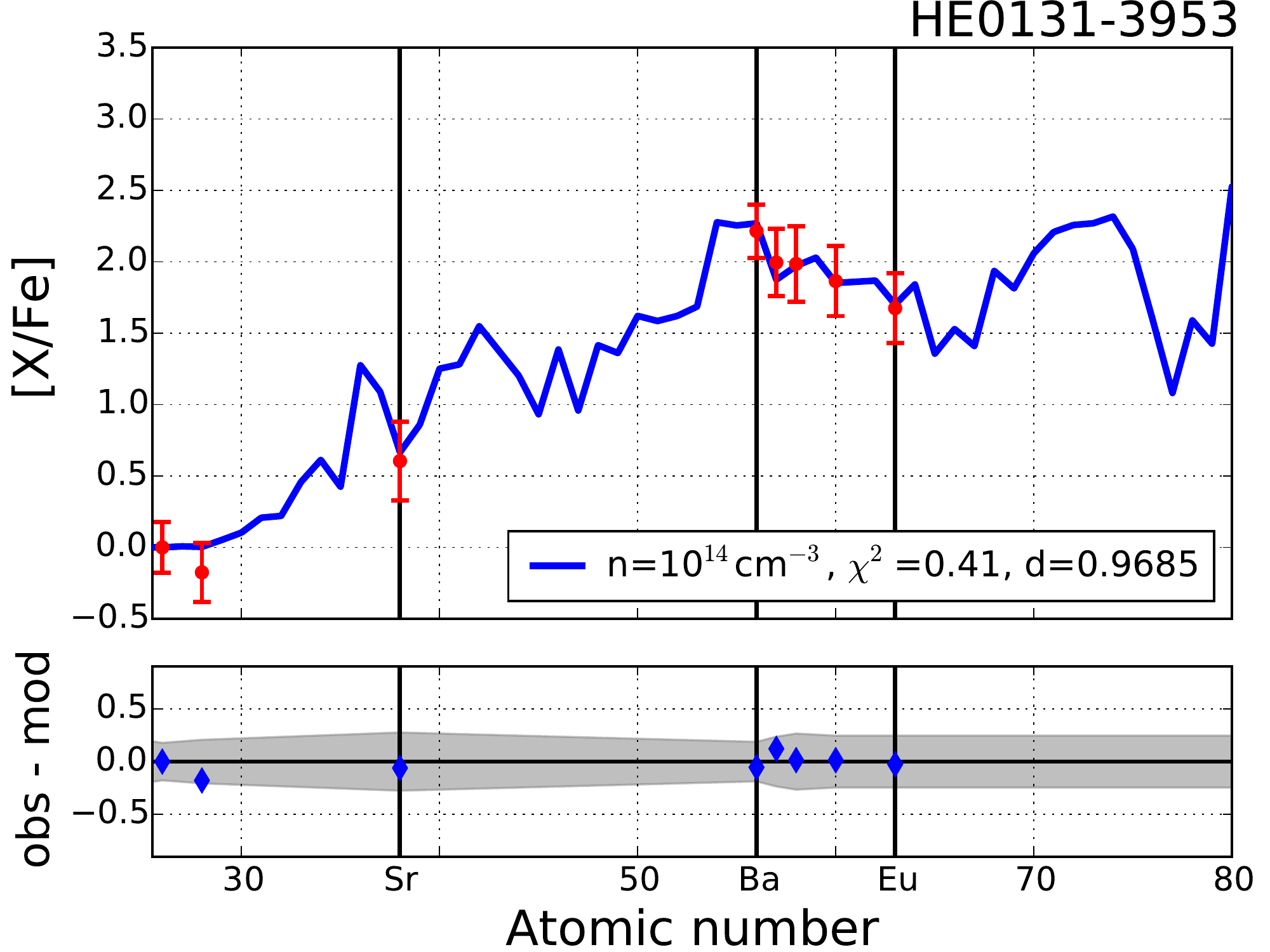}
\caption{Best fitting model for CEMP-$s$/$r$ star HE0131-3953. The best fitting s-process models with initial $r$-process enhancement can be found in Fig.~A8 from \citet{Bisterzo2012}.}
\end{minipage}
\hfill
\begin{minipage}[t]{.46\textwidth}
\vspace{12pt}
\centering
\includegraphics[ width=\linewidth]{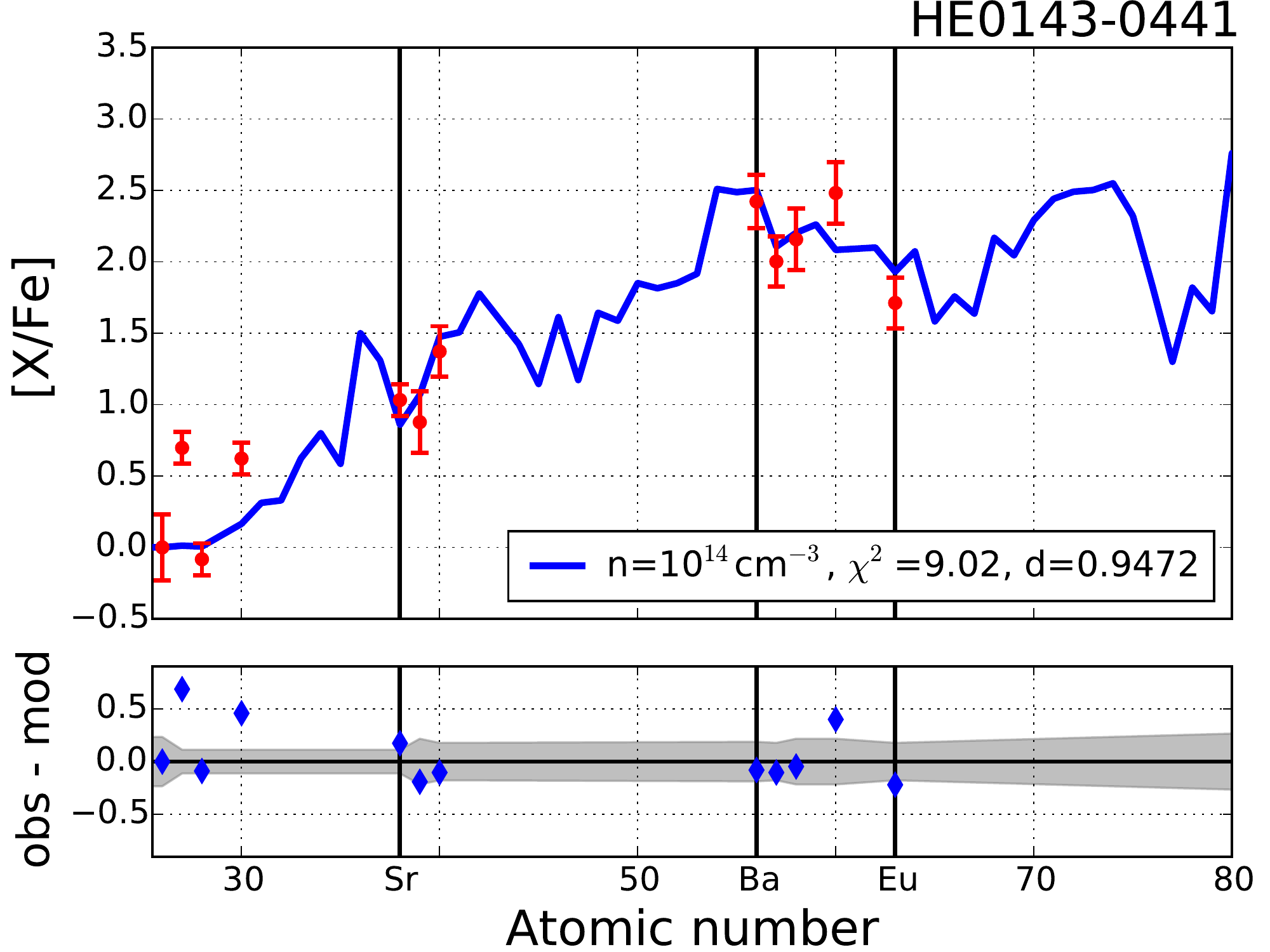}
\caption{Best fitting model for CEMP-$s$/$r$ star HE0143-0441. The best fitting s-process models with initial $r$-process enhancement can be found in Fig.~33 from \citet{Bisterzo2012}.}
\end{minipage}
\hfill
\begin{minipage}[t]{.46\textwidth}
\vspace{12pt}
\centering
\includegraphics[ width=\linewidth]{HE0338-3945-fit.pdf}
\caption{Best fitting model for CEMP-$s$/$r$ star HE0338-3945. The best fitting s-process models with initial $r$-process enhancement can be found in Fig.~19 from \citet{Bisterzo2012}.}
\end{minipage}
\hfill
\begin{minipage}[t]{.46\textwidth}
\vspace{12pt}
\centering
\includegraphics[ width=\linewidth]{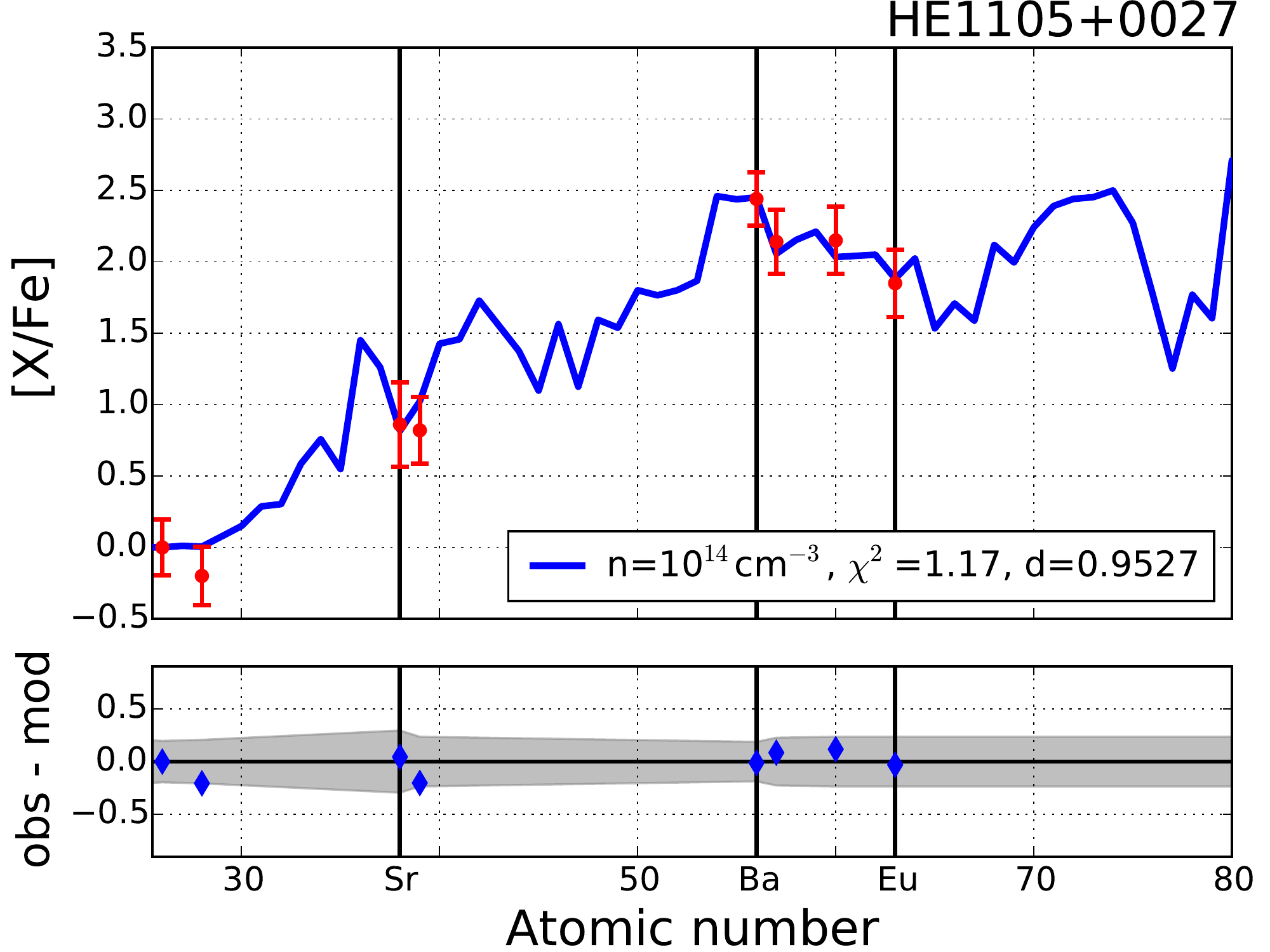}
\caption{Best fitting model for CEMP-$s$/$r$ star HE1105+0027. The best fitting s-process models with initial $r$-process enhancement can be found in Fig.~20 from \citet{Bisterzo2012}.}
\end{minipage}
\hfill
\begin{minipage}[t]{.46\textwidth}
\vspace{12pt}
\centering
\includegraphics[ width=\linewidth]{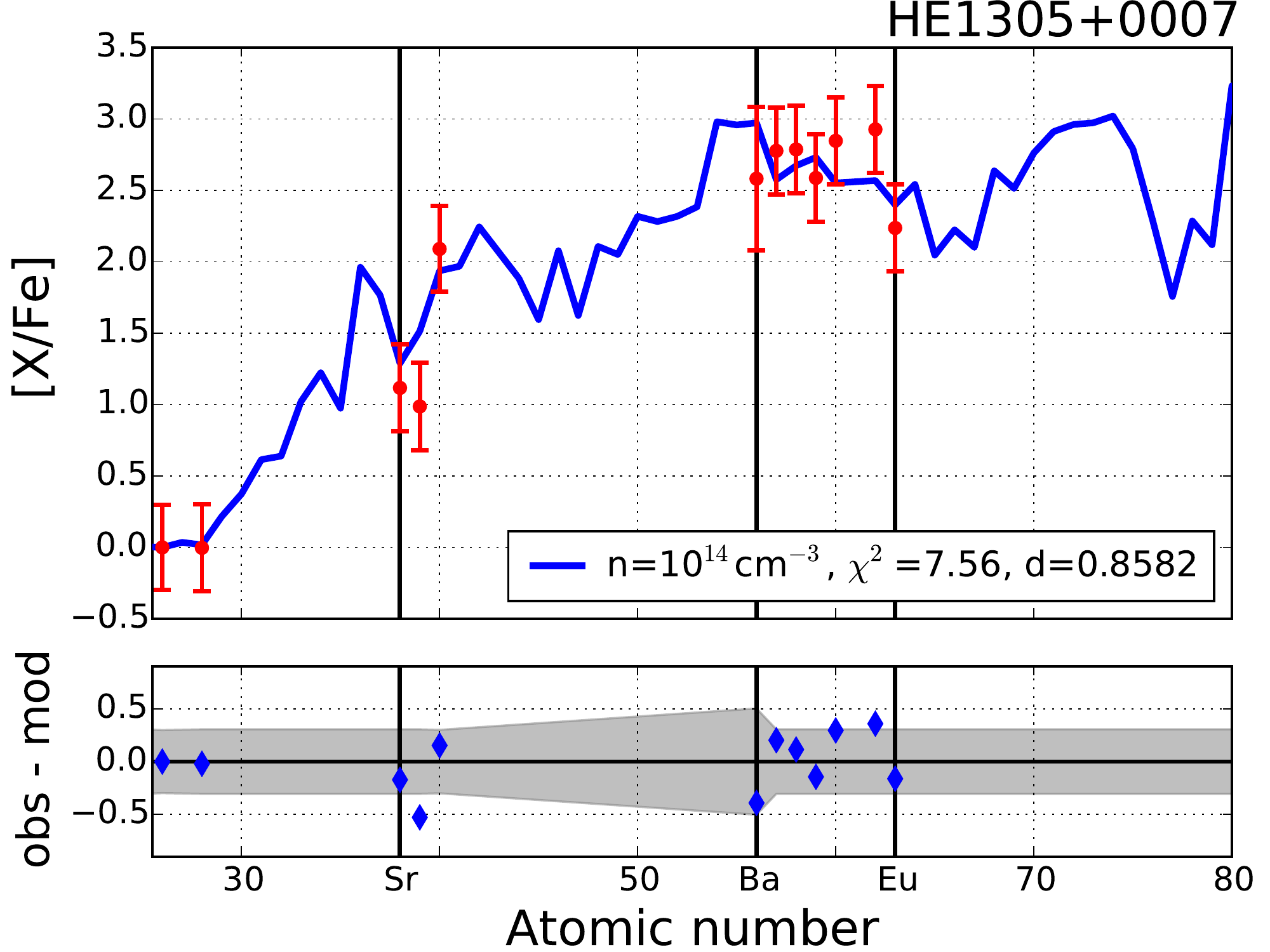}
\caption{Best fitting model for CEMP-$s$/$r$ star HE1305+0007. The best fitting s-process models with initial $r$-process enhancement can be found in Fig.~22 from \citet{Bisterzo2012}.}
\end{minipage}
\end{figure}
\begin{figure}
\begin{minipage}[t]{.46\textwidth}
\vspace{12pt}
\centering
\includegraphics[ width=\linewidth]{HE2148-1247-fit.pdf}
\caption{Best fitting model for CEMP-$s$/$r$ star HE2148-1247. The best fitting s-process models with initial $r$-process enhancement can be found in Fig.~21 from \citet{Bisterzo2012}.}
\end{minipage}
\hfill
\begin{minipage}[t]{.46\textwidth}
\vspace{12pt}
\centering
\includegraphics[ width=\linewidth]{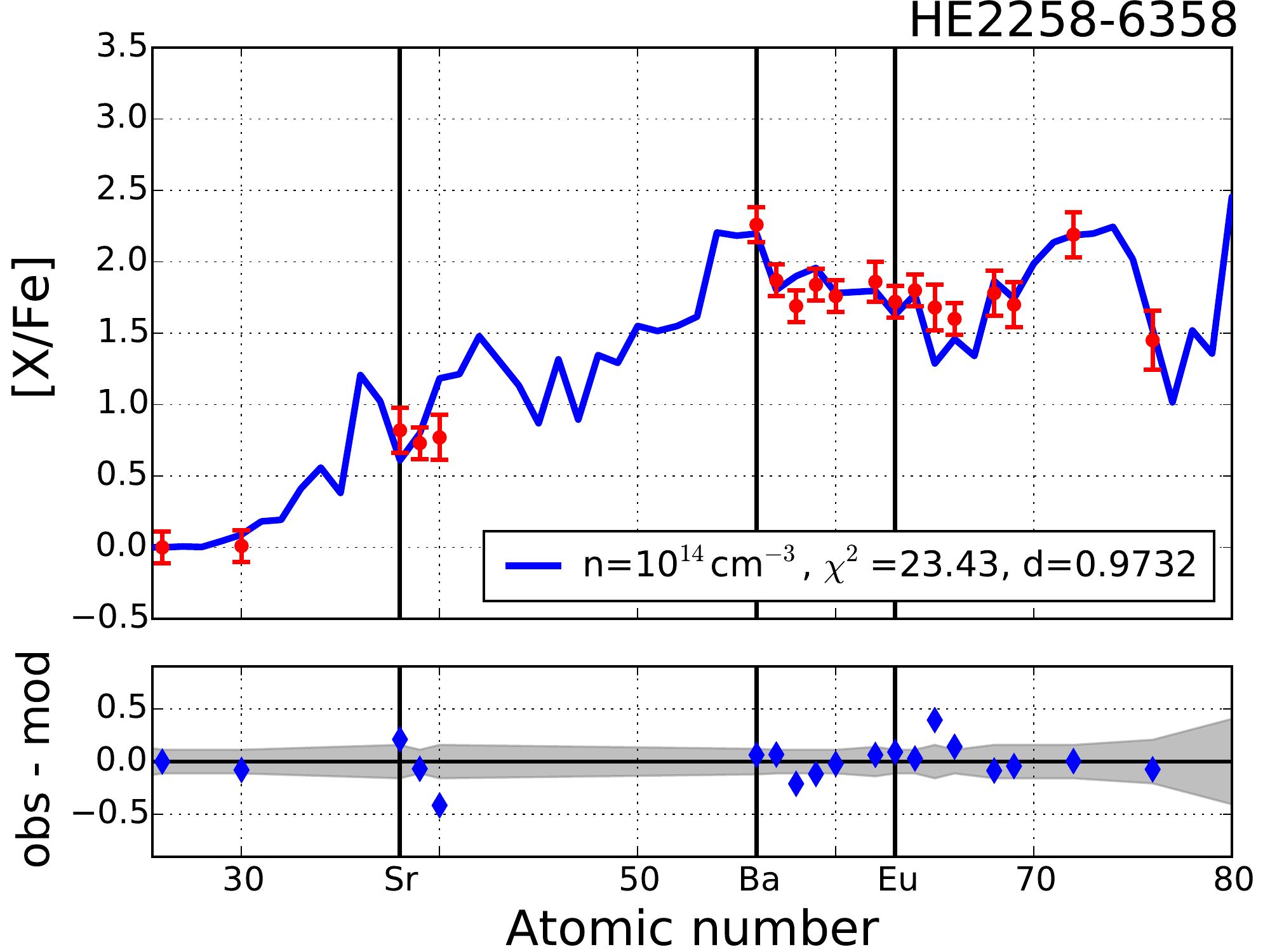}
\caption{Best fitting model for CEMP-$s$/$r$ star HE2258-6358.}
\end{minipage}
\hfill
\begin{minipage}[t]{.46\textwidth}
\vspace{12pt}
\centering
\includegraphics[ width=\linewidth]{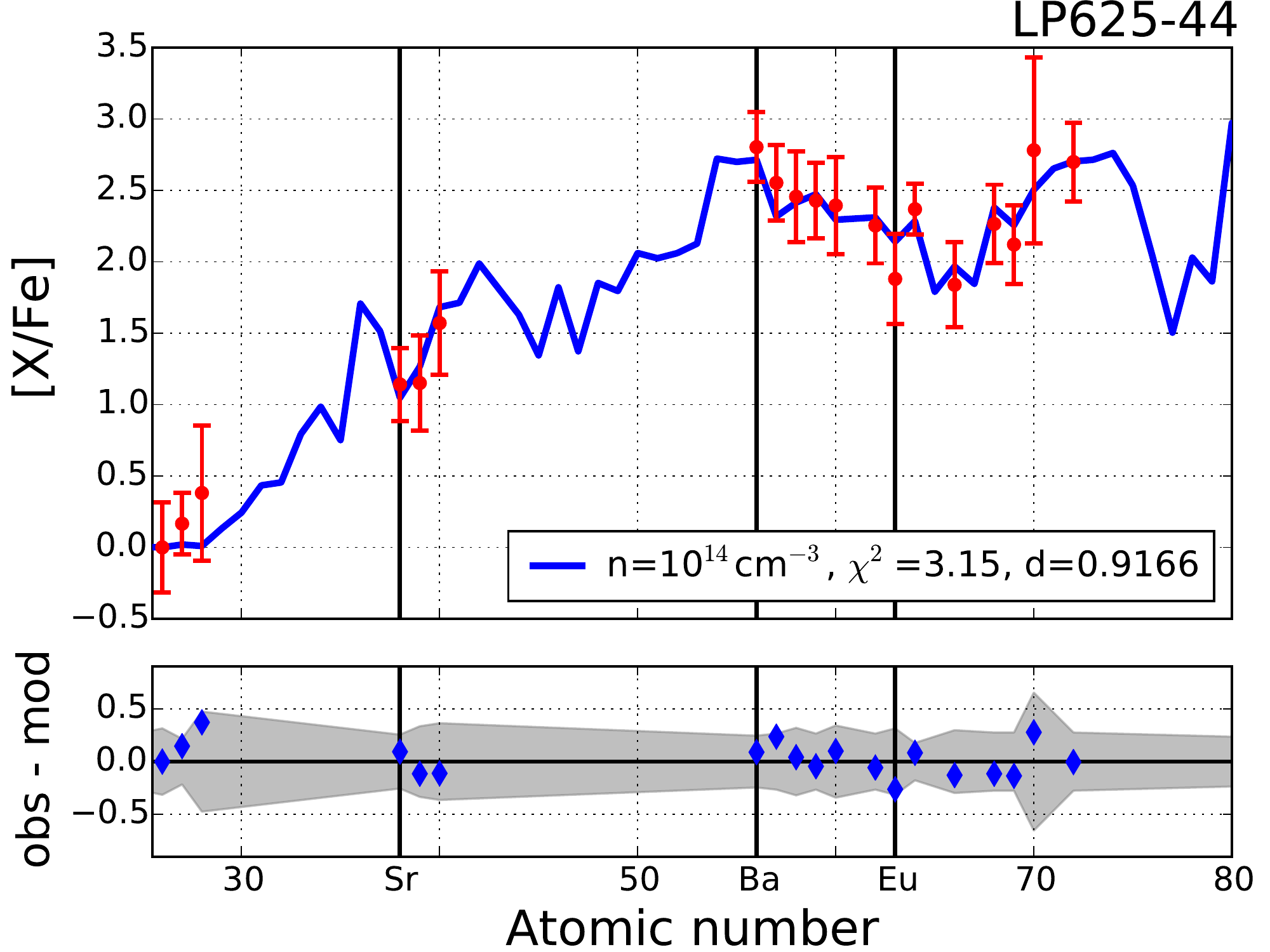}
\caption{Best fitting model for CEMP-$s$/$r$ star LP625-44. The best fitting s-process models with initial $r$-process enhancement can be found in Fig.~31 from \citet{Bisterzo2012} and with binary evolution in Fig.~A11 from \citet{Abate2015}.}
\end{minipage}
\hfill
\begin{minipage}[t]{.46\textwidth}
\vspace{12pt}
\centering
\includegraphics[ width=\linewidth]{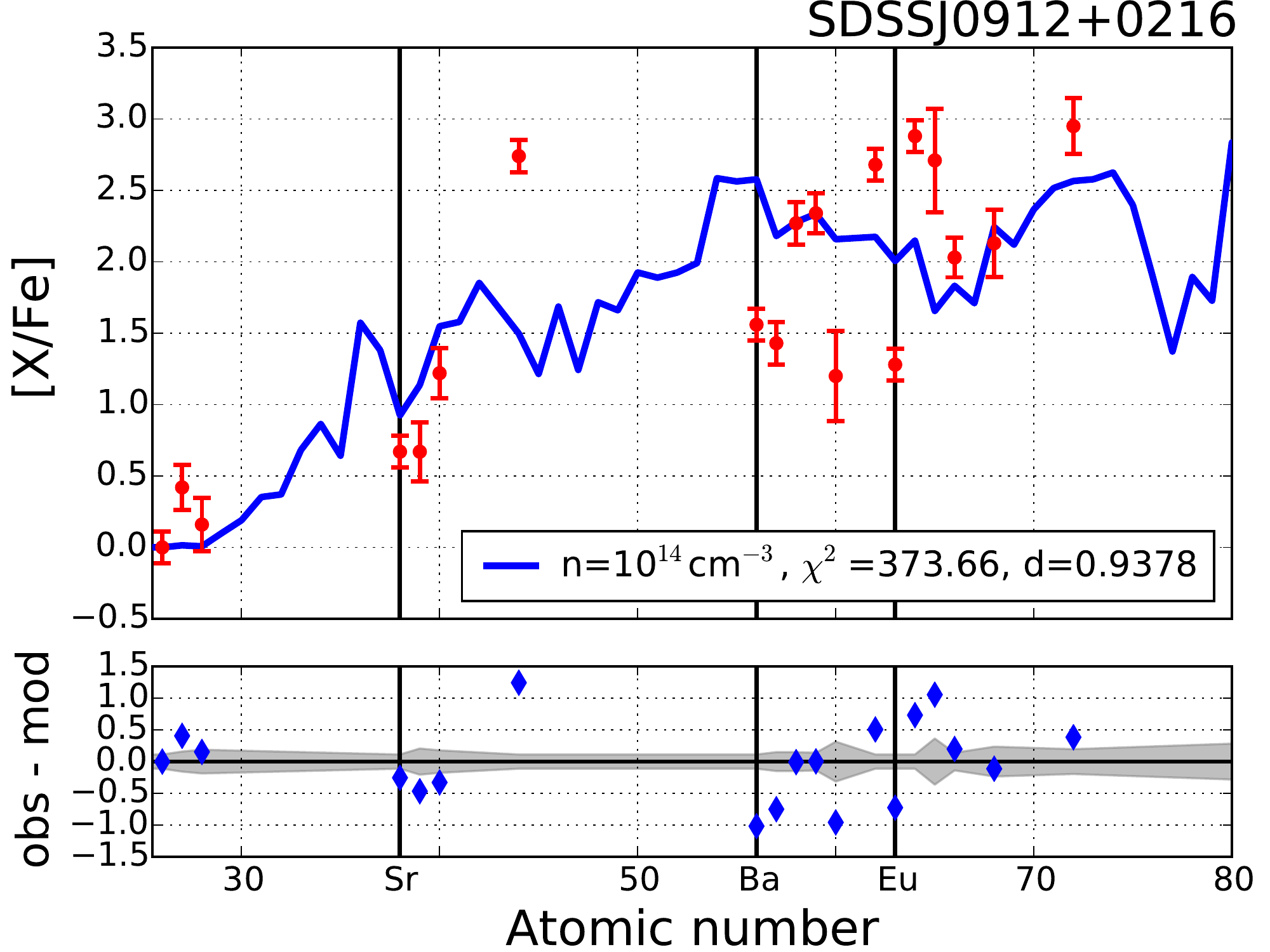}
\caption{Best fitting model for CEMP-$s$/$r$ star SDSSJ0912+0216. The best fitting s-process models with initial $r$-process enhancement can be found in Fig.~34 from \citet{Bisterzo2012}.}
\end{minipage}
\end{figure}

\end{document}